%% file: dynet_hd_v13_suppl.tex
\documentclass[twocolumn]{autart}
\usepackage[utf8]{inputenc}
\usepackage[T1]{fontenc}
\usepackage[scaled=0.95]{inconsolata}

\usepackage{latexsym, amsmath, amssymb, amsfonts,
            upgreek, mathrsfs, mathtools}
\usepackage{arydshln}   
\usepackage{empheq}     
\usepackage{enumitem}   

\usepackage{algorithm}
\usepackage{algpseudocode}

\usepackage{float, graphicx, caption, subcaption}
\usepackage{rotating}
\usepackage{multirow,booktabs}
\usepackage[usenames,dvipsnames]{xcolor}
\usepackage[colorlinks=true,linkcolor=magenta,citecolor=blue,
            urlcolor=cyan,filecolor=red]{hyperref}
\usepackage{tikz}

\input{supports/userdef-mathsym.tex}

\hypersetup{
  pdftitle={},
  pdfauthor={Zuogong YUE},
  pdfsubject={Automatica preprint},
  pdfcreator={Emacs version 25.1 + AUCTeX version 11.91}}

\begin{document}

\begin{frontmatter}

\title{Dynamic Network Reconstruction from \\ Heterogeneous Datasets \\
  {\small (Supplementary Version)}}


\author[Lux]{Zuogong Yue}\ead{zuogong.yue@uni.lu},
\author[Lux,SweH]{Johan Thunberg}\ead{johan.thunberg@uni.lu},
\author[UK]{Wei Pan}\ead{w.pan11@imperial.ac.uk},
\author[Swe]{Lennart Ljung}\ead{ljung@isy.liu.se},
\author[Lux]{Jorge Gon\c{c}alves}\ead{jorge.goncalves@uni.lu}

\address[Lux]{Luxembourg Centre for Systems Biomedicine, University of Luxembourg, L-4362, Luxembourg}
\address[UK]{Centre for Synthetic Biology and Innovation and Department of Bioengineering, Imperial College London, UK}
\address[SweH]{School of Information Technology, Halmstad University, SE-30118 Halmstad, Sweden}
\address[Swe]{Department of Electrical Engineering, Link\"oping University, Link\"oping, SE-58183, Sweden}


\begin{abstract}
  Performing multiple experiments is common when learning internal
  mechanisms of complex systems. These experiments can include
  perturbations to parameters or external disturbances. A challenging
  problem is to efficiently incorporate all collected data simultaneously
  to infer the underlying dynamic network. This paper addresses the
  reconstruction of dynamic networks from heterogeneous datasets under the
  assumption that underlying networks share the same Boolean structure
  across all experiments. Parametric models for dynamical structure
  functions are derived to describe causal interactions between measured
  variables. Multiple datasets are integrated into one regression problem
  with additional demands of group sparsity to assure network sparsity and
  structure consistency. To acquire structured group sparsity, we propose a
  sampling-based method, together with extended versions of $l_1$ methods
  and sparse Bayesian learning. The performance of the proposed methods is
  benchmarked in numerical simulation. In summary, this paper presents
  efficient methods on network reconstruction from multiple experiments,
  and reveals practical experience that could guide applications.
\end{abstract}

\end{frontmatter}


\section{Introduction}
\label{sec:introduction}

Network reconstruction has been widely applied in different fields to learn
interaction structures or dynamic behaviours, including systems biology,
computer vision, econometrics, social networks, etc.  With increasingly easy
access to time-series data, it is expected that networks can help to explain
dynamics, causal interactions and internal mechanisms of complex systems. For
instance, biologists use causal network inference to determine critical genes
that are responsible for diseases in pathology, e.g. \cite{Bar-Joseph2012}.

Boolean networks are introduced in \cite{Akutsu1999,Shmulevich2002} to
approximate the dynamics of an underlying processes by Markov chain models
with binary state-transition matrices.  However, it may fail to capture
complex dynamics.  Another popular model is Bayesian networks,
e.g. \cite{Murphy1999}. Although it delivers causality information,
Bayesian networks are defined as a type of \emph{directed acyclic graphs}
(see \cite[p.~127]{Pearl1988} for domains of probabilistic
models). Feedback loops are particularly important in applications. Even
though dynamic Bayesian network is able to handle loops in digraphs
\cite{Murphy2002}, it requires huge amounts of data (repetitive
measurements of each time point), which may not be practical in biology.
Regarding causality, \emph{Granger causality} (GC) is originally proposed
in \cite{Granger1969a} and defined in general based on conditional
independence (see e.g. \cite{Eichler2007b}).  However, as mentioned in
\cite{Granger1969a}, this approach may be ``problematic in deterministic
settings, especially in dynamic systems with weak to moderate coupling''
\cite{Sugihara2012}. With minor conditions on spectral density, GC can be
inferred by estimating vector autoregression models \cite{Hsiao1982}.  A
recent work \cite{Chiuso2012} proposes a kernel-based system identification
method with sparsity constraints to infer GC graphs.

There have been many applications of network reconstruction by identifying
simplified dynamical models, e.g. \cite{VanSomeren2000}, \cite{Friston2003}
and \cite{Beal2005}. However, identifiability remains untouched in these
study, which is essential to guarantee reliability of results. To study
network identifiability, the work in \cite{Goncalves2008} proposed
\emph{dynamical structure functions} (DSFs) for linear time-invariant (LTI)
systems as a general network representation. Similar network models
appeared in \cite{VandenHof2013,Weerts2015}, and their differences are
discussed in \cite{warnick2015}. Theoretical results on network
identifiability were firstly presented in \cite{Goncalves2008}. Successive
work in \cite{Hayden2014,Hayden2014a} studies network identifiability from
intrinsic noise and \cite{Hayden2016} presents identifiability results for
sparse networks. Moreover, there is a structural point of view for
topological identification of networks from data, e.g.,
\cite{Materassi2010,Materassi2012b}.

Many biological applications have replica to remove effects of noise.
However, only a handful of or less replica are typically available in
practice, meaning that averages may not be statistically significant. The
work in \cite{Pan2015} proposes an idea to integrate replica data as a
whole, which will be adopted in this paper. It considers a particular
reconstruction problem for nonlinear systems, which assumes full-state
measurements (no hidden states), basis functions known \emph{a priori} and
linearity in parameters. Our work will focus on general network
reconstruction problems of LTI systems, with existence of hidden states. In
practice, system parameters in multiple experiments can be fairly different
due to high variability of biological processes \cite{He2012}. On the other
hand, for identical organisms, the interconnection structure, e.g. genetic
regulation, should be identical over experimental repetitions. Using DSFs
as a general network model class, the method proposed in the paper allows
system parameters to be fairly different, while keeping the network
topology consistent across replica.

\begin{figure}[htb]   
  \centering
  \includegraphics[width=.48\textwidth]{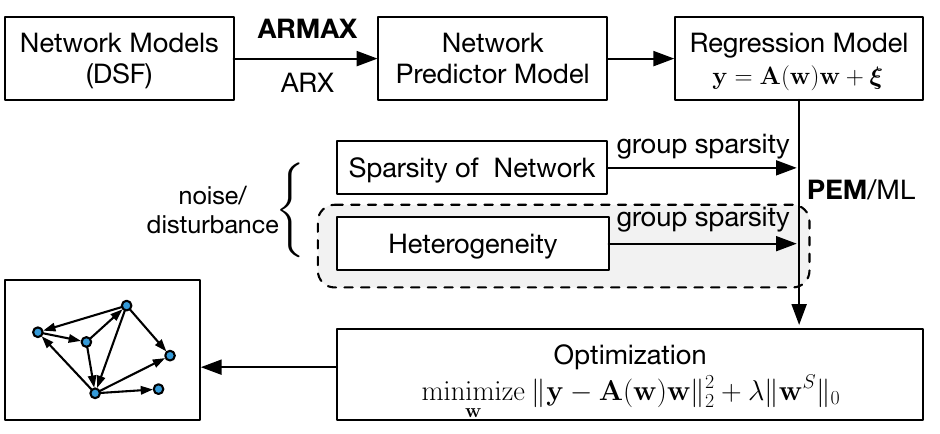}
  \caption{An overview of the network reconstruction method.}
  \label{fig:framework}
\end{figure}

The  paper is organised in modules, illustrated by
Fig.~\ref{fig:framework}: 1) derivation of regression problems for network
reconstruction (Section~3); 2) dealing with heterogeneity data (Section~4); 3)
solvers to the target optimisation problem (Section~5).  The paper is closed by
benchmark studies on random sparse network reconstruction. A supplement to
technical details and other illustrative examples are provided as appendices in
\cite{Yue2016a}. Codes in MATLAB and benchmark data are available at
\url{https://github.com/oracleyue/{dynet_dt-mat, datasets}}.

\section{Problem Description}
\label{sec:problem-description}

Let $Y \triangleq \{y(t), t \in \mathbb{Z}\}$, $U \triangleq \{u(t), t \in \mathbb{Z}\}$ be multivariate time series of dimension $p$ and $m$, respectively, where the elements could be deterministic ($y(t) \in \mathbb{R}^p$, $u(t) \in \mathbb{R}^m$) or be real-valued random vectors defined on probability spaces $(\Omega, \mathscr{F}, \mathbb{P})$. We usually assume that $u(t)$ is deterministic in practice, which is interpreted as controlled input signals.

\subsection{Linear dynamic networks}
\label{subsec:linear-dy-net}

Consider the network model for LTI systems, called \emph{dynamical structure function} (DSF) \cite{Goncalves2008,Chetty2015},
\begin{equation}
  \label{eq:dynet-model}
  y(t) = Q(q) y(t) + P(q) u(t) + H(q) e(t),
\end{equation}
where $y(t) = [y_1(t), \dots, y_p(t)]^T$,
$u(t) = [u_1(t), \dots, u_m(t)]^T$, a $p$-variate i.i.d.
$e(t) = [e_1(t), \dots, e_p(t)]^T$ with zero mean and identity covariance
matrix,
\begin{equation*}
  \begin{alignedat}{2}
    Q(q) &=[Q_{ij}(q)]_{p \times p}, \quad\quad && Q_{ii}(q) = 0,\ \forall i, \\
    P(q) &=[P_{ij}(q)]_{p \times m}, \quad\quad && H(q) =[H_{ij}(q)]_{p
      \times p}
  \end{alignedat}
\end{equation*}
$Q_{ij}(q), P_{ij}(q), H_{ij}(q)$ are single-input-single-output (SISO)
real-rational transfer functions, and $q$ is the forward-shift operator,
i.e. $q y(t) = y(t+1)$, $q^{-1} y(t) = y(t-1)$. Here $Q_{ij}(q)$
($i \neq j$) are strictly proper, and $P_{ij}, H_{ii}(q)$ are proper. See
Appendix~\ref{appdix:dsf} in \cite{Yue2016a} for more on DSFs. The
networks represented by DSFs can be defined by extending the
capacity functions of standard networks in the graph theory.
  Let $\mathcal{G} = (V,E)$ be a digraph, where the vertex set
  $V = \{y_1, \dots, y_p, u_1, \dots, u_m, e_1, \dots, e_p\}$, and
  the directed edge set $E$ is defined by
  \begin{displaymath}
    \hspace*{-25mm}
    \begin{array}{@{}ll@{\;}l@{\;\,}l@{\;\,}l@{\;}l}
      \text{ $\centerdot$} &(y_j, y_i) &\in E &\Leftrightarrow &Q_{ij}(q) &\neq 0,\\
      \text{ $\centerdot$} &(u_k, y_i) &\in E &\Leftrightarrow &P_{ik}(q) &\neq 0,\\
      \text{ $\centerdot$} &(e_l, y_i) &\in E &\Leftrightarrow &H_{il}(q) &\neq 0,\\
      \text{ $\centerdot$} &(y_i, u_k) &\notin E, &\multicolumn{3}{@{\,}l}{(y_i, e_l) \notin E,}
    \end{array}
  \end{displaymath}
  for all $i,j,k$ and $l$. And let the \emph{capacity function} $f$ be a
  map defined as
  \begin{equation*}
    \begin{array}{rr@{\;\,}c@{\;\,}l}
      f: & E & \rightarrow & {S}_{\text{TF}} \\
         & (y_j, y_i) & \mapsto & Q_{ij}(q) \;\;\text{or}\;\; (u_k, y_i)
                                  \mapsto  P_{ik}(q) \;\;\text{or}\;\; \\
         &(e_l, y_i) &\mapsto& H_{il}(q),
    \end{array}
  \end{equation*}
  where ${S}_{\text{TF}}$ is a subset of single-input-single-output (SISO)
  proper rational transfer functions. We call the tuple
  $\mathcal{N} \coloneqq (\mathcal{G}, f)$ a (linear) \emph{dynamic
    network}, and $\mathcal{G}$ the \emph{underlying digraph} of
  $\mathcal{N}$, which is also called (linear) \emph{Boolean dynamic
    network}.



To guarantee network identifiability, i.e. unique determination of DSFs from
input-output transfer functions, we need to assume either $H(q)$ or $P(q)$ being
square and diagonal, e.g. see \cite{Goncalves2008,Hayden2016a}. Here we take the
diagonal $H$ as an example,
and we further assume all assumptions in \cite{Hayden2016a} are satisfied.
\begin{assum}
  \label{assump:diag-H}
  The matrix $H$ is square, diagonal and full rank.
\end{assum}
Alternatively, we could use general $H$ but restrict $P$ to be square and
diagonal, and the forthcoming method can be easily modified correspondingly.  A
square and diagonal $P$ presents a practical way in experiments to guarantee
network identifiability, i.e. perturbing each measured variable.  Indeed, the
DSF with square diagonal $H$ is fairly special, which can be equivalently
rewritten such that reconstruction can be treated as identification of
input-output transfer functions with sparse input selection. However, this paper
uses a flexible framework that is valid for diagonal $P$ or $H$, or even
other network identifiability conditions.


\subsection{Network reconstruction from multiple experiments}
\label{subsec:net-infer-multi-exper}

Consider multiple experiments, where $\{Y^{[l]}, U^{[l]}\}_{l=1,\dots,L}$ denote
measurements from $L$ experiments. Let $\mathcal{N}((Q,P,H))$ be the dynamic
network $\mathcal{N}$ determined by $(Q,P,H)$, and $\mathcal{G}((Q,P,H))$ be the
corresponding Boolean dynamic network. The governing model
\eqref{eq:dynet-model} could be different in each experiment, denoted by
$(Q,P,H)^{[l]}, l = 1,\dots,L$.  In addition, $\mathcal{N}^0$ denotes a fixed
dynamic network, and $\mathcal{G}^0$ a fixed Boolean dynamic network.  The
datasets $\{Y^{[l]}, U^{[l]}\}_{l=1,\dots,L}$ are called \emph{homogeneous}, if
$\mathcal{N}( (Q,P,H)^{[l]} ) \equiv \mathcal{N}^0, \forall l$, i.e. the
measurements in multiple experiments are from the same dynamic network. And the
datasets $\{Y^{[l]}, U^{[l]}\}_{l=1,\dots,L}$ are said to be
\emph{heterogeneous}, if
$\mathcal{G}( (Q,P,H)^{[l]} ) \equiv \mathcal{G}_0, \forall l$ but
$\mathcal{N}( (Q,P,H)^{[l]} )$ are different between certain
$l \in \{1,\dots,L\}$.  The constraint
$\mathcal{G}( (Q,P,H)^{[l]} ) \equiv \mathcal{G}_0$ presents the common feature
shared by systems in multiple experiments.  For example, even though biological
individuals are different in nature, if they are all healthy or subject to the
same gene-level operations, it is fair to assume they have the same genetic
regularisation.
\begin{assum}
  \label{assump:main-multi-exp}
  The underlying systems in multiple experiments, which provide
  $\{Y^{[l]}, U^{[l]}\}_{l=1,\dots,L}$, satisfy that
  $\mathcal{G}( (Q,P,H)^{[l]} ) \equiv \mathcal{G}_0$ for any $l = 1,\dots,L$.
\end{assum}

The problem is to develop methods to infer the common Boolean network using the
datasets from multiple experiments satisfying
Assumption~\ref{assump:main-multi-exp}. In particular, we focus on the
heterogeneous case. To help to understand the problem and the operations in
Section~\ref{sec:net-pred-models} and \ref{sec:heterogen-data}, we construct a
specific example in Appendix~\ref{appdix:example} of \cite{Yue2016a}.

\section{Network model structures}
\label{sec:net-pred-models}

This section shows a standard procedure to parametrize network models and derive
corresponding regression problems.  It can be applied to other types of
parametric models, such FIR, ARARX, ARARMAX, etc. We use ARMAX as an example in
this paper.

\subsection{Element-wise ARMAX parametrisation}
\label{subsec:armax-model-struct}

Consider the network \emph{model description} of \eqref{eq:dynet-model} for system identification
\begin{equation}
  \label{eq:dynet-model-description}
  y(t) = Q(q, \theta) y(t) + P(q, \theta) u(t) + H(q, \theta) e(t),
\end{equation}
where $\theta$ is the model parameter. Its element-wise form is
\begin{equation}
  \label{eq:dynet-model-description-ith}
  \begin{array}{@{}l@{\:}l}
    y_i(t) = &\sum_{j=1}^p Q_{ij}(q, \theta) y_j(t) + \sum_{k=1}^m P_{ik}(q, \theta)
               u_k(t)\\
             &+ H_{ii}(q, \theta) e_i(t).
  \end{array}
\end{equation}
This is an MISO (multiple-input-single-output) transfer function in system
identification. We use the ARMAX model to parametrise
\eqref{eq:dynet-model-description-ith}, yielding that
\begin{equation*}
  \begin{array}{@{}l@{\:}l}
    A_i(q) y_i(t) = &\sum_{j=1, j\neq i}^p B^y_{ij}(q) y_j(t) + \\
                    &\sum_{k=1}^m B^u_{ik}(q) u_k(t) + C_{ii}(q) e_i(t)
  \end{array} \quad \text{with}
\end{equation*}
\begin{equation*}
  \begin{array}{rcr@{\!}r@{\,}rr@{\,}l}
    A_i(q)     &= &1 + &a_{i1} &q^{-1} + \cdots +& a_{i n_i^a} &q^{- n_i^a}, \\
    B^y_{ij}(q) &= & &b^y_{ij1} &q^{-1} + \cdots +& b^y_{ij n_{ij}^{by}} &q^{- n_{ij}^{by}}, \\
    B^u_{ij}(q) &= & &b^u_{ij1} &q^{-1} + \cdots +& b^u_{ij n_{ij}^{bu}} &q^{- n_{ij}^{bu}}, \\
    C_i(q)     &= & 1 + &c_{i1}&q^{-1} + \cdots +& c_{i n_i^c} &q^{- n_i^c}.
  \end{array}
\end{equation*}
Hence the ARMAX model for \eqref{eq:dynet-model-description} is
\begin{equation}
  \label{eq:armax-model}
  A(q) y(t) = B^y(q) y(t) + B^u(q) u(t) + C(q) e(t),
\end{equation}
where $A = \diag(A_1, \dots, A_p)$, $C = \diag(C_1, \dots, C_p)$,
$B^y = \big[ B_{ij}^y \big]_{p \times p}$ with zero diagonal, and
$B^u = \big[ B_{ij}^u \big]_{p \times p}$.
It is easy to see the following relations
\begin{equation}
  \label{eq:dynet-vs-armax}
  Q(q,\theta) = A^{-1} B^y, \; P(q,\theta) = A^{-1} B^u, \; H(q,\theta) = A^{-1}C.
\end{equation}


\subsection{Predictor model and regression forms}
\label{subsec:regression-forms}

Considering network model \eqref{eq:dynet-model} and noticing that $[I-Q(q)]$ is invertible, we have
\begin{math}
  y(t) = (I- Q)^{-1} P u(t) + (I-Q)^{-1} H e(t)
       \triangleq G_u u(t) + G_e e(t).
\end{math}
We refer to \cite[pp.~70]{Ljung1998} for the one-step-ahead prediction of $y$,
\begin{math}
  \hat{y}(t | t-1) = G_e^{-1} G_u u(t) + (I - G_e^{-1} ) y(t),
\end{math}
and thus the network \emph{predictor model} of \eqref{eq:dynet-model-description} is given by
\begin{math}
    \hat{y}(t | t-1)  = H^{-1} P u(t) +\, H^{-1} (Q + H - I) y(t).
\end{math}
The one-step-ahead predictor of the ARMAX model follows by substituting \eqref{eq:dynet-vs-armax}
\begin{equation}
  \label{eq:dynet-armax-predictor-model}
  \hat{y}(t | \theta)  = C^{-1} B^u u(t) + (C^{-1}B^y + I - C^{-1}A) y(t),
\end{equation}
where $\hat{y}(t|t-1) \triangleq \hat{y}(t|\theta)$ to emphasize the dependency on model parameter $\theta$.


Rewriting \eqref{eq:dynet-armax-predictor-model} and adding
$[I - C(q)] \hat{y}(t|\theta)$ on both sides, it yields that
\begin{equation}
  \label{eq:dynet-armax-predictor-model-2}
  \begin{array}{ll}
    \hat{y}(t|\theta) = &B^u(q) u(t) + \big[ B^y(q) - \big( A(q) - I \big) \big] y(t) \\
                        &+ \big(C(q) - I \big) \big[ y(t) - \hat{y}(t|\theta) \big].
  \end{array}
\end{equation}
To formulate a regression form, let us introduce the prediction error
\begin{math}
  \varepsilon(t|\theta) \coloneqq y(t) - \hat{y}(t|\theta),
\end{math}
and consider the prediction of the $i$-th output $y_i(t)$
\begin{equation}
  \label{eq:dynet-armax-predictor-model-ith}
  \begin{array}{ll}
    \hat{y}_i(t|\theta) =
    &\bar{B}^u_i(q) u(t) + \big[ \bar{B}^y_i(q) -
      \big( \bar{A}_i(q) - \bar{I}_i \big) \big] y(t) \\
    &+ \big({C}_i(q) - \bar{I}_i \big)
      \big[ y_i(t) - \hat{y}_i(t|\theta) \big],
  \end{array}
\end{equation}
where $\bar{A}_i, \bar{B}^y_i, \bar{B}^u_i$, $\bar{I}_i$ are the corresponding
$i$-th rows of $A, B^y, B^u$ and $I$.  Using notations
\begin{equation}
  \label{eq:regression-form-phi}
  \begin{array}{l@{\,}r@{\;\,}l@{\,}r@{}l@{\:}l@{\:}l@{}l@{\,}l}
    \varphi(t, \theta_i) \triangleq \big[
    & y_1(t-1)  &\dots  &y_1&(t&-&n_{i1}^{by}&)
    &\dots  \\
    & -y_i(t-1)  &\dots &-y_i&(t&-&n_i^a&)
    &\dots \\
    & y_p(t-1)  &\dots  &y_p&(t&-&n_{ip}^{by}&) \\
    & u_1(t-1)  &\dots  &u_1&(t&-&n_{i1}^{bu}&)
    &\dots \\
    & u_i(t-1)  &\dots  &u_i&(t&-&n_{ii}^{bu}&)
    &\dots \\
    & u_m(t-1)  &\dots  &u_m&(t&-&n_{im}^{bu}&) \\
    & \varepsilon_i(t-1)  &\dots  &\varepsilon_i&(t&-&n_i^c&)\phantom{,} &\big]^T,
  \end{array}
\end{equation}
\begin{equation}
  \label{eq:regression-form-theta}
  \def\arraystretch{.8}
  \begin{array}{@{\!}r@{\:}lc@{\,}rllc@{\,}rclc@{\,}rl}
    \cdashline{2-4} \cdashline{6-8}
    \theta_i \triangleq \big[
    &\multicolumn{1}{:l@{\,}}{b^y_{i11}} &\cdots &\multicolumn{1}{r@{\,}:}{b^y_{i1 n_{i1}^{by}}} &\;\cdots\;
    &\multicolumn{1}{:l@{\,}}{a_{i1}} &\cdots &\multicolumn{1}{r@{\,}:}{a_{i n_i^a}} &\;\cdots \\
    \cdashline{2-4} \cdashline{6-8} \\
    \cdashline{2-4}
    &\multicolumn{1}{:l@{\,}}{b^y_{ip1}} &\cdots &\multicolumn{1}{r@{\,}:}{b^y_{ip n_{ip}^{by}}} &\\
    \cdashline{2-4} \\
    \cdashline{2-4} \cdashline{6-8}
    &  \multicolumn{1}{:l@{\,}}{b^u_{i11}} &\cdots &\multicolumn{1}{r@{\,}:}{b^u_{i1 n_{i1}^{bu}}} &\;\cdots
    &\multicolumn{1}{:l@{\,}}{b^u_{ii1}} &\cdots &\multicolumn{1}{r@{\,}:}{b^u_{ii n_{ii}^{bu}}}  &\;\cdots \\
    \cdashline{2-4} \cdashline{6-8} \\
    \cdashline{2-4}
    &\multicolumn{1}{:l@{\,}}{b^u_{im1}} &\cdots &\multicolumn{1}{r@{\,}:}{b^u_{im n_{im}^{bu}}}  &\\
    \cdashline{2-4} \\
    \cdashline{2-4}
    &\multicolumn{1}{:l@{\,}}{c_{i1}} &\cdots &\multicolumn{1}{r@{\,}:}{c_{i n_i^c}} &\multicolumn{1}{l@{\!}}{\big]^T,} & \multicolumn{3}{l}{\text{($M$ blocks)}} &&&& & \\
    \cdashline{2-4}\\
  \end{array}
\end{equation}
where $M = p+m+1$ (if ARX, $M= p+m$), we obtain a pseudo-linear regression form
\begin{equation}
  \label{eq:pseudolinear-regression-form}
  \hat{y}_i(t|\theta_i) = \varphi^T(t, \theta_i) \theta_i,\quad i=1,\dots,p.
\end{equation}

\begin{rem}
  \label{rmk:comments-block-structure-w}
  Note that there is an important relation between the framed parameter blocks
  in \eqref{eq:regression-form-theta} and the network structure.  Each arc in
  the digraph corresponds to a transfer function from input $u_j$ or $y_j$ to
  output $y_i$. The parameters of the transfer function are given in the block
  with parameters $b^u_{ij\cdot}$ or $b^y_{ij\cdot}$ together with
  $a_{i\cdot}$. This relation will be used in Section~\ref{sec:heterogen-data}
  and be illustrated by Fig.~\ref{fig:eg-w}.
\end{rem}

\section{Heterogeneous datasets}
\label{sec:heterogen-data}

This section starts dealing with heterogeneous datasets, which will be
integrated with additional constraints to guarantee
Assumption~\ref{assump:main-multi-exp}.

\subsection{Regression forms of multiple datasets}
\label{subsec:regr-model-form}

Consider the regression problems \eqref{eq:pseudolinear-regression-form} in
network reconstruction, where the $p$ problems can be treated independently.
Therefore, without loss of generality, it is assumed in later sections that we
are dealing with \eqref{eq:pseudolinear-regression-form} for the $i$-th output
variable $y_i$ by default.  To avoid the lengthy index notations in
\eqref{eq:regression-form-phi} and \eqref{eq:regression-form-theta}, which are
unnecessary in later discussions, we encapsulate them by introducing the
following notations
\begin{equation}
  \label{eq:pseudolinear-regression-form-samples}
  \begin{array}{r@{\:}l}
    {\mathbf{y}}^{[l]} &\triangleq
    \big[\:
      {y}_i(t_1|\theta_i) \; \cdots \; {y}_i(t_{N_l}|\theta_i)
    \:\big]^T, \\
    \mathbf{A}^{[l]}(\mathbf{w}^{[l]}) &\triangleq
    \big[\:
      \varphi(t_1,\theta_i) \; \cdots \; \varphi(t_{N_l},\theta_i)
    \:\big]^T,
  \end{array}
\end{equation}
where $\mathbf{w}^{[l]} \triangleq \theta_i$; $y_i, \varphi$ and $\theta_i$
correspond to the $l$-th experiment; and \eqref{eq:pseudolinear-regression-form}
is evaluated at $\{t_1, \dots, t_{N_l}\}$.  We then have the following
expression
\begin{equation}
  \label{eq:regression-form-exp-dataset}
    \mathbf{y}^{[l]} = \mathbf{A}^{[l]}(\mathbf{w}^{[l]})\, \mathbf{w}^{[l]} + \pmb{\xi}^{[l]}, \quad l = 1,\dots, L,
\end{equation}
where
\begin{equation}
  \label{eq:regression-form-exp-parameters}
  \begin{aligned}
    \mathbf{A}^{[l]} &\triangleq
    \begin{bmatrix}
      \mathbf{A}_{:,1}^{[l]}\;\; & \mathbf{A}_{:,2}^{[l]}\;\; & \cdots\;\; & \mathbf{A}_{:,M}^{[l]}
    \end{bmatrix},\\ 
    \mathbf{w}^{[l]} & \hspace*{.6pt}
    \begin{array}[t]{l@{}lllll}
      \triangleq \Big[
        &\big(\mathbf{w}_1^{[l]}\big)^T &\cdots &\big(\mathbf{w}_i^{[l]}\big)^T &\cdots &\big(\mathbf{w}_p^{[l]}\big)^T, \\
        &\big(\mathbf{w}_{p+1}^{[l]}\big)^T &\cdots &\big(\mathbf{w}_{p+i}^{[l]}\big)^T &\cdots &\big(\mathbf{w}_{p+m}^{[l]}\big)^T, \\
        &\big(\mathbf{w}_M^{[l]}\big)^T \Big]^T
    \end{array}\\
    \pmb{\xi}^{[l]} &\triangleq
    \begin{bmatrix}
      \xi^{[l]}(t_1)\;\; & \xi^{[l]}(t_2)\;\; & \cdots\;\; & \xi^{[l]}(t_{N_l})
    \end{bmatrix},
  \end{aligned}
\end{equation}
$\mathbf{w}^{[l]}$ is partitioned into $M$ blocks as illustrated in
\eqref{eq:regression-form-theta}, $\mathbf{A}^{[l]}_{:,j}$ ($j=1,\dots,M$)
denotes the blocks of $\mathbf{A}^{[l]}$ that is partitioned correspondingly as
$\mathbf{w}^{[l]}$, and $\pmb{\xi}^{[l]}$ denotes the prediction error, which
represents the part of the output $\mathbf{y}^{[l]}$ that cannot be predicted
from past data using the chosen model classes.
Letting
\begin{equation}
  \label{eq:wkc-wk-def}
  \begin{array}{l@{\:}l}
    \mathbf{w}_k &\triangleq
    \left[
    \begin{array}{ccccc}
      \big(\mathbf{w}_k^{[1]}\big)^T & \big| & \cdots
      & \big|  & \big(\mathbf{w}_k^{[L]}\big)^T
    \end{array}
    \right]^T,\\[5pt]
    \mathbf{w} &= \left[
    \begin{array}{ccccc}
      \mathbf{w}_1^T &  \big| & \cdots & \big| &  \mathbf{w}_M^T
    \end{array}
    \right]^T,
  \end{array}
\end{equation}
we integrate all datasets by stacking \eqref{eq:regression-form-exp-dataset} for
each dataset and rearranging blocks of matrices, yielding
\eqref{eq:regression-form-exp-all}, and, for simplicity, use
$\mathbf{y} = \mathbf{A}(\mathbf{w}) \mathbf{w} + \pmb{\xi}$ to denote
\eqref{eq:regression-form-exp-all-N-blocks}. One may see a specific example in
Appendix~\ref{appdix:example} of \cite{Yue2016a} to help to understand how the
arrangement is made.

\begin{figure*}[htp]
  \centering

  \begin{subequations}
    \label{eq:regression-form-exp-all}
    \begin{align}
      \label{eq:regression-form-exp-all-C-blocks}
      \left[
        \begin{array}{c}
          \mathbf{y}^{[1]}\\
          \vdots \\
          \mathbf{y}^{[L]}
        \end{array}
      \right]
      & =
      \underbrace{
        \left[
        \begin{array}{ccc|c|ccc} \mathbf{A}^{[1]}_{:,1}(\mathbf{w}^{[1]})
          & \ldots & \mathbf{A}^{[1]}_{:,M}(\mathbf{w}^{[1]}) &  & & &   \\
          &  & & \ddots  & &  &\\
          & &  & & \mathbf{A}^{[L]}_{:,1}(\mathbf{w}^{[L]}) & \ldots
            & \mathbf{A} ^{[L]}_{:,M}(\mathbf{w}^{[L]})
          \end{array}
        \right]
      }_{C~\textbf{Blocks}}
      \left[
        \begin{array}{c}
          \mathbf{w}^{[1]}\\\hline
          \vdots \\ \hline
          \mathbf{w}^{[L]}
        \end{array}
      \right]
      +
      \left[
        \begin{array}{c}
          \pmb{\xi}^{[1]}\\\hline
          \vdots \\\hline
          \pmb{\xi}^{[L]}
        \end{array}
      \right]\\
      \label{eq:regression-form-exp-all-N-blocks}
      &=
      \underbrace{
        \left[
          \begin{array}{ccc|c|ccc}
            \mathbf{A}^{[1]}_{:,1}(\mathbf{w}^{[1]})& &  &
            &\mathbf{A}^{[1]}_{:,M}(\mathbf{w}^{[1]})& &\\
            & \ddots & & \cdots  & & \ddots &\\
            & & \mathbf{A} ^{[L]}_{:,1}(\mathbf{w}^{[L]})& &   &
            & \mathbf{A}^{[L]}_{:,M}(\mathbf{w}^{[L]})
          \end{array}
        \right]
      }_{M ~\textbf{Blocks}}
      \left[
        \begin{array}{c}
          \mathbf{w}_1 \\ \hline \vdots \\ \hline \mathbf{w}_M
        \end{array}
      \right]
      +
      \left[
        \begin{array}{c}
          \pmb{\xi}^{[1]}\\\hline
          \vdots \\ \hline
          \pmb{\xi}^{[L]}
        \end{array}
      \right]
    \end{align}
  \end{subequations}
\end{figure*}


\subsection{Simultaneous sparsity regularization}
\label{subsec:simult-spars-regul}

Now we consider the two essential requirements for network reconstruction from heterogeneous datasets:
1) sparse networks is acquired in the presence of noise;
2) $\mathbf{w}^{[l]}$ is required to give the same network topology for all $l$,
i.e. the inference results $(\hat{Q},\hat{P},\hat{H})^{[l]}$ satisfy
$\mathcal{G}((\hat{Q},\hat{P},\hat{H})^{[l]}) \equiv \mathcal{G}^0$ for any $l$ (see Section~\ref{subsec:net-infer-multi-exper}).
\begin{figure}[htb]   
  \centering
  \includegraphics[width=.45\textwidth]{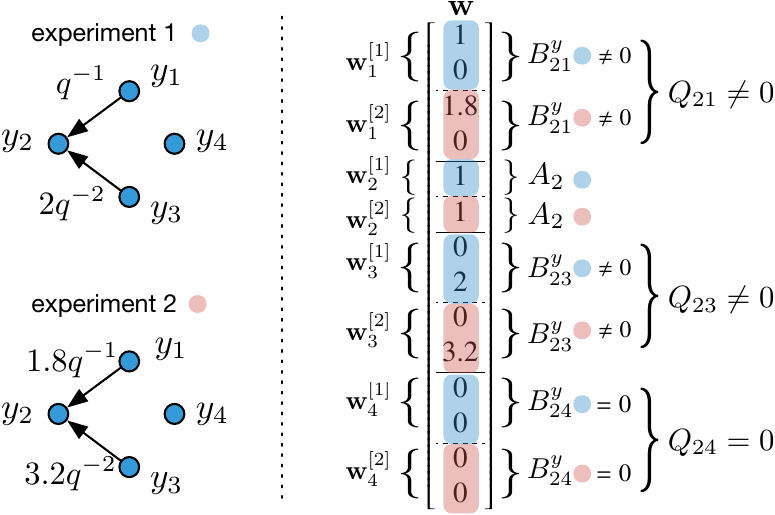}
  \caption{An example of $\mathbf{w}$ in the setup of multiple experiments.}
  \label{fig:eg-w}
\end{figure}

Let us introduce a term of group sparsity based on $\mathbf{w}$,
\begin{equation}
  \label{eq:group-sparsity-2types}
  \mathbf{w}^S \coloneqq
  \big[
    \|\mathbf{w}_1\|_2,\;   \cdots,\;   \|\mathbf{w}_M\|_2
  \big]^T,
\end{equation}
where $\|\cdot\|_2$ denotes the $l_2$-norm of vectors and $M$ is the number of
groups/blocks in \eqref{eq:regression-form-exp-parameters}.
The dimensions of each block $\mathbf{w}_k^{[l]}$ and $\mathbf{w}_k$ in
\eqref{eq:wkc-wk-def} are saved in two vectors $\rho^{E}$ and $\rho^{S}$, respectively
\begin{equation}
  \label{eq:block-dim-rho}
  \begin{array}{ll}
    &\rho \triangleq \big[ n_{i1}^{by}, \dots, n_i^a, \dots, n_{ip}^{by}, n_{i1}^{bu}, \dots,
      n_{ip}^{bu}, n_i^c \big]^T,\\
    &\rho^{E} \coloneqq \rho \otimes \mathbf{1}_L, \qquad \rho^{S} \coloneqq L \rho,
  \end{array}
\end{equation}
where the elements of $\rho$ are defined with respect to
\eqref{eq:regression-form-theta}, $\mathbf{1}_L$ is a $L$-dimensional column
vector of $1$'s, and the index $i$ in $\rho$ indicates the regression problem
for $y_i$, as assumed by default.

Group sparsity is demanded to guarantee networks sparsity and that the
network structure is consistent over replica (i.e. the interconnection
structure determined by the $\mathbf{w}^{[l]}$'s are identical for all
$l$).  The sparsity is imposed on each large group $\mathbf{w}_k$,
$k = 1, \dots, M$, and the penalty term is $\lambda \|\mathbf{w}^{S}\|_0$,
where $\lambda \in \mathbb{R}^+$. The mechanism on how group sparsity
functions is described as follows.

Recall that the setup \eqref{eq:regression-form-exp-all} allows the system
parameters to be different in values for different $l$'s.  Note that each
small block $\mathbf{w}_k^{[l]}$ corresponds to an arc in the underlying
digraph of the dynamical system in the $l$-th experiment (e.g., see
Fig.~\ref{fig:eg-w}).  The $\|\mathbf{w}_k\|_2$ chosen to be zero yields
that all $\mathbf{w}_k^{[l]}, l= 1,\dots,L$ are equal to zeros. It implies
that the arc corresponding to $\mathbf{w}_k^{[l]}$ does not exist in
dynamic networks for any $l$.  In addition, thanks to the effect of noise,
it is nearly guaranteed that $\mathbf{w}_k^{[l]}$ is not identical to zero
for almost all $l$ if $\|\mathbf{w}_k\|_2 \neq 0$.  This is how the group
sparsity (defined via $\mathbf{w}^S$) guarantees that the resultant
networks of different datasets share the same topology.  Moreover, when a
classical least squares objective is augmented with a penalty term of
$\lambda \|\mathbf{w}^S\|_0$, the optimal solution favors zeros of
$\mathbf{w}_k, k = 1,\dots, M$, which guarantees the sparsity of network
structures.
In summary, to perform network reconstruction from heterogeneous datasets, we solve the following optimization problem
\begin{equation}
  \label{eq:regression-zero-regularization}
  \minimize{\mathbf{w}} \|\mathbf{y} - \mathbf{A}(\mathbf{w}) \mathbf{w}\|^2 + \lambda \|\mathbf{w}^{S}\|_0,
\end{equation}
where the norm $\|\cdot\|$ can be simply the $l_2$-norm in the classic
$l_1$ methods (see Section~\ref{sec:treat-l1-methods}), or the $l_2$-norm
weighted by inverse noise covariance $\Sigma^{-1}$ of $\pmb{\xi}$ in
\eqref{eq:regression-form-exp-all}, as specified in details later in
Section~\ref{sec:sparse-bayes-learn} and \ref{sec:mcmc-methods-arx}.


\section{Methods for structured group sparsity}
\label{sec:methods-group-sparse}

The problem of network reconstruction from heterogeneous data has been
tackled in Section~\ref{sec:net-pred-models} and \ref{sec:heterogen-data},
which results in an regularised optimisation problem. The onus is to solve
the group sparsity problem \eqref{eq:regression-zero-regularization}
provided with the fixed partition given in \eqref{eq:wkc-wk-def}. However,
subject to the progress of regularised optimisation, this section focuses
on handling the ARX, or any other parametrization that results in
\eqref{eq:regression-zero-regularization} being a regularised
linear least square problem. Indeed, there could be heuristic methods to
solve the general \eqref{eq:regression-zero-regularization}, due to lack of
rigorousness in mathematics, which will not be covered in this paper. We
organise this section as a separate module such that any solver that
readers trust could be integrated with Section~\ref{sec:net-pred-models}
and \ref{sec:heterogen-data} to solve reconstruction problems from multiple
experiments.

\subsection{Classic $l_1$ methods}
\label{sec:treat-l1-methods}

Two well-known methods are firstly introduced and extended to solve the
``linear case'' of \eqref{eq:regression-zero-regularization}, specifically,
classic $l_1$ heuristic methods and sparse Bayesian learning (SBL). The
classic $l_1$ methods refer to the family of methods that use best convex
approximation of \eqref{eq:regression-zero-regularization}. Due to the
general sparsity structure of $\mathbf{w}^S$, the resultant problem will
have a more general form than the standard group LASSO in \cite{Simon2012}.
The extended formulations of $l_1$ methods for structured group sparsity
will be presented, together with an ADMM algorithm for large-scale problems.

\subsubsection{Convex approximation}
\label{subsec:a-basic-case-arx-model}

As addressed in Section~\ref{subsec:regression-forms}, choosing ARX to
parametrize network models results in a linear regression, in which $\mathbf{A}$
does not depend on $\mathbf{w}$ in \eqref{eq:regression-form-exp-all}.
The treatment of classical group LASSO yields
\begin{equation}
  \label{eq:regression-l1-regularization}
  \minimize{\mathbf{w}} \|\mathbf{y} - \mathbf{A} \mathbf{w}\|_2^2 + \lambda \|\mathbf{w}^{S}\|_1,
\end{equation}
where
\begin{equation}
  \label{eq:group-lasso-relaxation}
  \|\mathbf{w}^{S}\|_1 = \sum_{i=1}^M \sqrt{\rho_i^{S}} \| \mathbf{w}_i \|_2,
\end{equation}
and $\lambda \in \mathbb{N}_+$.  This is a convex optimization and has been
soundly studied (e.g., \cite{Yuan2006}).
To achieve a better approximation of the $l_0$-norm, alternatively one may use
\emph{iterative reweighted $l_1/l_2$ methods} (e.g. see
\cite{Candes2008,Chartrand2008}).  Applying to group sparsity, both
methods turn to a similar scheme (differing in the usage of $\|\cdot\|_2$ or
$\|\cdot\|^2_2 $ for blocks of $\mathbf{w}$ in
\eqref{eq:iter-reweight-l1-method} and
\eqref{eq:iter-reweight-l1-method-choosing-weights}).
Here we present the solution using the $l_1$ method,
\begin{equation}
  \label{eq:iter-reweight-l1-method}
    \mathbf{w}^{(k+1)} = \argmini{\mathbf{w}}
    \|\mathbf{y} - \mathbf{Aw}\|_2^2 +
    \lambda \sum_{i=1}^M \nu_i^{(k)}  \sqrt{\rho_{i}^{S}} \| \mathbf{w}_i \|_2,
\end{equation}
where
\begin{equation}
  \label{eq:iter-reweight-l1-method-choosing-weights}
  \nu_i^{(k)} = \Big[ \| \mathbf{w}_i^{(k)} \|_2
  + \epsilon^{(k)} \Big]^{-1}
\end{equation}
and $k$ is the index of iterations.  In regard to the selection of $\epsilon$,
$\{\epsilon^{(k)}\}_{k=1,2,\dots}$ should be a sequence converging to zero, as
addressed in \cite{Chartrand2008} based on the \emph{Unique Representation
  Property}. It suggests in \cite{Chartrand2008} a fairly simple update rule of
$\epsilon$, i.e. $\epsilon^{(k)} \in (0,1)$ is reduced by a factor of 10 until
reaching a minimum of $10^{-8}$ (the factor and lower bound could be tuned
specifically). One may also adopt an adaptive rule of $\epsilon$ given
in \cite{Candes2008}.

\subsubsection{ADMM for large-scale problems}
\label{subsec:prox-algor-admm}

To solve the convex optimization in Section~\ref{subsec:a-basic-case-arx-model},
for example, \emph{CVX} for MATLAB could be an easy solution.  However, the
computation time could be enormous for large-dimension problems. This section
presents algorithms using \emph{proximal methods} and \emph{ADMM}
(\cite{Parikh2013}) to handle large-dimension network reconstruction.

Let us first consider \eqref{eq:regression-l1-regularization},
which is rewritten as
\begin{equation}
  \label{eq:group-lasso-relax-fg}
  \minimize{\mathbf{w}} f(\mathbf{w}) + g(\mathbf{w}),
\end{equation}
where $f(\mathbf{w}) \triangleq (1/2) \|\mathbf{y} - \mathbf{A} \mathbf{w}\|_2^2$,
$g(\mathbf{w}) \triangleq \lambda \|\mathbf{w}^{S}\|_1$, $\lambda$ is twice larger than the value in \eqref{eq:group-lasso-relaxation}.
Given $\nabla f(\mathbf{w}) = \mathbf{A}^T (\mathbf{A} \mathbf{w} - \mathbf{y})$, the \emph{proximal gradient method} is to update $\mathbf{w}$ by
$\mathbf{w}^{(k+1)} = \mathbf{prox}_{\gamma g}(\mathbf{w}^{(k)} - \gamma \nabla
f(\mathbf{w}^{(k)})),\, \gamma \in \mathbb{R}_+$, where $k$ is the iteration index
and $\mathbf{prox}_f(\cdot)$ denotes the standard \emph{proximal operator} of function $f$ (see \cite{Parikh2013,Boyd2011}).
It is easy to see that $g(\mathbf{w}) = \sum_{i=1}^N g_i(\mathbf{w}_i)$, where $g_i(\mathbf{w}_i) \triangleq \lambda \sqrt{\rho_i^S} \|\mathbf{w}_i\|_2$. Firstly we partition the variable $\mathbf{v}$ of $\mathbf{prox}_{\gamma g}(\mathbf{v})$ in the same way as $\mathbf{w}$ in terms of $\mathbf{w}_i, i =1,\dots,M$, i.e. $\mathbf{v} = [\mathbf{v}_1^T,\dots, \mathbf{v}_M^T]^T$.
Then we calculate the proximal operator $\mathbf{prox}_{\gamma g_i}(\mathbf{v}_i)$, which equals
\begin{equation}
  \label{eq:prox-gk}
  \mathbf{prox}_{\gamma g_i}(\mathbf{v}_i) = \Big( 1 - \gamma \lambda
    \sqrt{\rho_i^S}/\|\mathbf{v}_i\|_2 \Big)_+ \mathbf{v}_i,
\end{equation}
where $(\cdot)_+$ replaces each negative elements with 0.  It follows that
\begin{equation}
  \label{eq:prox-g}
  \mathbf{prox}_{\gamma g}(\mathbf{v}) =
  \begin{bmatrix}
    \left(\mathbf{prox}_{\gamma g_1}(\mathbf{v}_1) \right)^T &
    \cdots &
    \left(\mathbf{prox}_{\gamma g_M}(\mathbf{v}_M) \right)^T
  \end{bmatrix}^T.
\end{equation}
The value of $\gamma$ needs to be selected appropriately so as to guarantee the
convergence. One simple solution is using line search methods, e.g., see
Section~4.2 in \cite{Parikh2013}.

Provided with the above calculations, it is straightforward to implement the
\emph{(accelerated) proximal gradient method} (see \cite[chap.~4.3]{Parikh2013}). To implement ADMM, the proximal
operator of $f(\mathbf{w})$ needs to be calculated,
\begin{equation}
  \label{eq:prox-f}
  \mathbf{prox}_{\gamma f}(\mathbf{v}) = (I + \gamma \mathbf{A}^T \mathbf{A})^{-1} (\gamma \mathbf{A}^T\mathbf{y} + \mathbf{v}).
\end{equation}
Given $\mathbf{prox}_{\gamma g}(\mathbf{v})$ as \eqref{eq:prox-gk} and
\eqref{eq:prox-g}, the ADMM method is presented in Algorithm~\ref{alg:admm-method}.

\begin{algorithm}
  \caption{ADMM method}
  \label{alg:admm-method}
  \begin{algorithmic}[1]
    \State Precompute $\mathbf{A}^T \mathbf{A}$ and $\mathbf{A}^T \mathbf{y}$

    \State \textbf{given} an initial value $\mathbf{w}^0, \mathbf{z}^0,
    \mathbf{u}^0$, $\gamma^0 = 1$, and $\beta = 1/2$

    \Repeat
    \State $\gamma \gets \gamma^{(k)}$
    \Repeat
    \State $\hat{\mathbf{w}} \gets \mathbf{prox}_{\gamma f}(\mathbf{z}^{(k)} - \mathbf{u}^{(k)})$ using \eqref{eq:prox-f}
    \State \textbf{break} if $f(\hat{\mathbf{w}}) \leq f(\mathbf{w}^{(k)}) + \nabla f(\mathbf{w}^{(k)})^T (\hat{\mathbf{w}} - \mathbf{w}^{(k)}) + (1/2\gamma) \|\hat{\mathbf{w}} - \mathbf{w}^{(k)}\|_2^2$
    \State $\gamma \gets \beta \gamma$
    \Until{;}
    \State $\mathbf{w}^{(k+1)} \gets \hat{\mathbf{w}}$, $\gamma^{(k+1)} \gets \gamma$
    \State Compute $\mathbf{prox}_{\gamma g_i}(\mathbf{w}^{(k+1)}_i + \mathbf{u}^{(k)}_i)$
    by \eqref{eq:prox-gk} for $i=1,...,M$
    \State $\mathbf{z}^{(k+1)} \gets \mathbf{prox}_{\gamma g}(\mathbf{w}^{(k+1)} + \mathbf{u}^{(k)})$ using \eqref{eq:prox-g}
    \State $\mathbf{u}^{(k+1)} \gets \mathbf{u}^{(k)} + \mathbf{w}^{(k+1)} - \mathbf{z}^{(k+1)}$
    \Until{any standard stopping criteria}
  \end{algorithmic}
\end{algorithm}

To use this algorithm for the iterative reweighted $l_1$ method
\eqref{eq:iter-reweight-l1-method}, we only need to modify \eqref{eq:prox-gk}, which now
should be
\begin{equation}
  \label{eq:prox-gk-iter-l1}
  \mathbf{prox}_{\gamma g_i}(\mathbf{v}_i) = \Big( 1 - \gamma \lambda \nu_i \sqrt{\rho_i^S}/\|\mathbf{v}_i\|_2 \Big)_+ \mathbf{v}_i.
\end{equation}
In each ``outer'' loop indicated by \eqref{eq:iter-reweight-l1-method}, we update
$\nu_i$ by \eqref{eq:iter-reweight-l1-method-choosing-weights} and implement
ADMM as Algorithm~\ref{alg:admm-method} to solve \eqref{eq:iter-reweight-l1-method}.

\subsection{Sparse Bayesian learning}
\label{sec:sparse-bayes-learn}

Another method is \emph{sparse Bayesian learning}, which was originally
proposed in \cite{Tipping2001} to heuristically acquire sparsity. Its
effectiveness and performance are further analysed in \cite{Wipf2004} and
compared to the iterative reweighted $l_1/l_2$ methods in theory in
\cite{Wipf2010}. Our problem demands a more general form of group sparsity
and noise covariance structures than that specified in \cite{Wipf2007}. It
includes how to structure priors to match the specification of
$\mathbf{w}^S$ and an extended EM algorithm for generalised noise
covariance structures.

Consider
\begin{equation}
  \label{eq:regression-model-bayes}
  \mathbf{y} = \mathbf{A} \mathbf{w} + \pmb{\xi}, \quad \pmb{\xi} \sim \mathcal{N}(0, \Sigma),
\end{equation}
where $\Sigma = \diag(\sigma_1^2I_1, \dots, \sigma_L^2I_L)$ with
$\sigma_l \in \mathbb{R}_+$; identity matrix $I_l$ is of dimension $N_l$
(that denotes the length of time series from the $l$-th experiment);
$\mathbf{y}$ is of dimension $M_\mathbf{y}$; and $\mathbf{w}$ is of dimension
$M_\mathbf{w}$ (it is easy to see $M_\mathbf{y} = \sum_{l=1}^L N_l$ and
$N_\mathbf{w} = \|\rho^{S}\|_1$ from \eqref{eq:regression-form-exp-dataset} and
\eqref{eq:group-sparsity-2types}).  Most SBL papers (e.g.,
\cite{Tipping2001,Wipf2010,Wipf2007,Wipf2004}) consider a simple form of noise
variance $\Sigma = \sigma^2 I$, which may fail to be a fair approximation in our
study if noise variances in different experiments are significantly
different.
This section will extend SBL to handle different group sizes and diagonal noise
variance matrices.

\subsubsection{Model prior formulation}
\label{subsec:model-prior-form}

Suppose the \emph{automatic relevance determination} (ARD) prior \cite{Tipping2001,Wipf2004} in SBL is given as
\begin{equation}
  \label{eq:prior}
  p(\mathbf{w}; \Gamma) = \frac{1}{(2\pi)^{M_\mathbf{w}/2} |\Gamma|^{1/2}} \exp\left(-\frac{1}{2}\mathbf{w}^T \Gamma^{-1} \mathbf{w}\right).
\end{equation}
In regard to hyperparameter $\Gamma$, we will enforce a specific structure
empirically to impose group sparsity.  For each $\mathbf{w}_j$, $j=1,\dots,M$,
  \begin{equation}
    \label{eq:prior-w-type2}
    p(\mathbf{w}_j; \gamma_j) = \mathcal{N}(0, \gamma_j I_j), \;\;
    p(\mathbf{w}; \pmb{\gamma}) = \prod_{j=1}^{M} p(\mathbf{w}_j; \gamma_j),
  \end{equation}
  where  $\gamma_j \in \mathbb{R}_+$, $I_j$ is a $LM_j \!\times\! LM_j$
  identity matrix,
  \begin{equation}
    \label{eq:def-Nk}
    M_j =
    \begin{cases}
      n_{ij}^{by} & \text{if } 1 \leq j \leq p \text{ and } j \neq i, \\
      n_i^a      & \text{if } j = i,\\
      n_{ij}^{bu} & \text{if } p+1 \leq j \leq p+m = M, \\
    \end{cases}
  \end{equation}
  the index $i$ in (\ref{eq:def-Nk}) indicates that we are dealing with the
  $i$-th output (see
  (\ref{eq:regression-form-theta}),
  (\ref{eq:pseudolinear-regression-form-samples})),
  and $\pmb{\gamma} \triangleq [\gamma_1, \dots, \gamma_M ]^T$.  Therefore,
  $\Gamma$ is constructed as
\begin{equation}
  \label{eq:def-Gammas}
    \Gamma = \diag\big( \gamma_1 I_1, \cdots, \gamma_M I_M \big).
\end{equation}
For simplicity, we will interchangeably use two notations for the prior
\begin{math}
  p(\mathbf{w}; \Gamma)  \triangleq p(\mathbf{w}; \pmb{\gamma}).
\end{math}

\subsubsection{Parameter estimation}
\label{subsec:hyperp-estim}

The likelihood function of \eqref{eq:regression-model-bayes} is Gaussian,
\begin{equation}
  \label{eq:bayes-likelihood}
  p(\mathbf{y} \mid \mathbf{w}; \Sigma) =
  \left(2\pi \right)^{-\frac{M_\mathbf{y}}{2}} |\Sigma|^{-\frac{1}{2}} \exp \left( \|\mathbf{y}-\mathbf{A}\mathbf{w}\|_{\Sigma^{-1}}^2 \right).
\end{equation}
The hyperparameters $\pmb{\gamma}$, along with the error variance $\Sigma$ can
be estimated from data by \emph{evidence maximisation} (or called \emph{type-II
  maximum likelihood}), i.e.  marginalising over the weights $\mathbf{w}$ and
then perform maximum likelihood. The marginalised
probabilistic density function can be analytically calculated
\begin{equation}
  \label{eq:marginalized-pdf}
  \begin{array}{ll}
    p(\mathbf{y}; \pmb{\gamma}, \Sigma)
    &\displaystyle = \int p(\mathbf{y}\mid\mathbf{w}; \Sigma) p(\mathbf{w}; \pmb{\gamma}) \mathrm{d} \mathbf{w}\\
    &\displaystyle = \frac{1}{(2\pi)^{M_\mathbf{y}/2} |\Sigma_{\mathbf{y}}|^{1/2}} \exp\left(-\frac{1}{2} \mathbf{y}^T \Sigma_\mathbf{y}^{-1} \mathbf{y}\right),
  \end{array}
\end{equation}
where $\Sigma_{\mathbf{y}} \triangleq \Sigma + \mathbf{A} \Gamma
\mathbf{A}^T$.
With fixed values of the hyperparameters, the posterior density is Gaussian, i.e.
\begin{equation}
  \label{eq:bayes-posterir}
  p(\mathbf{w} \mid \mathbf{y}; \pmb{\gamma}, \Sigma) = \mathcal{N}(\pmb{\mu}, \Sigma_{\mathbf{w}})
\end{equation}
with $\pmb{\mu} = \Sigma_{\mathbf{w}} \mathbf{A}^T \Sigma^{-1} \mathbf{y}$ and
$\Sigma_{\mathbf{w}} = \left( \mathbf{A}^T \Sigma^{-1} \mathbf{A} + \Gamma^{-1}
\right)^{-1}$. Once we have $\pmb{\gamma}$ and $\Sigma$ estimated via type-II maximum likelihood, we can choose our weights $\hat{\mathbf{w}}$ via
\begin{equation}
  \label{eq:map-estimate-w}
  \hat{\mathbf{w}} = \pmb{\mu} = \left( \mathbf{A}^T \Sigma_{\mathrm{ML}}^{-1} \mathbf{A} + \Gamma_\mathrm{ML}^{-1} \right)^{-1} \mathbf{A}^T \Sigma^{-1}_{\mathrm{ML}}\mathbf{y}.
\end{equation}

\subsubsection{Algorithms}
\label{subsec:byes-algorithms}

There are several ways to compute ${\Gamma}_\mathrm{ML}$ and
$\Sigma_\mathrm{ML}$: \emph{expectation maximization} (EM) in \cite{Wipf2004},
fixed-point methods in \cite{Tipping2001}, and \emph{difference of convex
  program} (DCP) in \cite{Pan2015}. Here we introduce the EM approach and the others
can be similarly derived.

The EM method proceeds by treating the weights $\mathbf{w}$ as hidden variables and then maximizing
\begin{equation*}
  \mathbb{E}_{\mathbf{w} | \mathbf{y}; \pmb{\gamma}, \Sigma} \,\big[p(\mathbf{y}, \mathbf{w}; \pmb{\gamma}, \Sigma)\big],
\end{equation*}
where $p(\mathbf{y}, \mathbf{w}; \pmb{\gamma}, \Sigma) = p(\mathbf{y} |
\mathbf{w}; \Sigma) p(\mathbf{w}; \pmb{\gamma})$ is the likelihood of the
complete data $\{\mathbf{w}, \mathbf{y}\}$.
The whole EM algorithm for the extended SBL is summarized as follows: given the
estimates $\pmb{\gamma}^{(k)}$ and $\Sigma^{(k)}$, at the
($k\!+\!1$)-th iterate,
\begin{itemize}
\item E-step:
  \begin{math}
    \mathbb{E}_{\mathbf{w}| \mathbf{y}; \pmb{\gamma}^{(k)}, \Sigma^{(k)}}
    (\mathbf{w}_i^2)= \left( \Sigma_\mathbf{w} \right)_{i,i} + \pmb{\mu}_i^2.
  \end{math}
\item M-step:
  \vspace*{-1em}
  \begin{equation*}
    \hspace*{-6mm}
    \begin{array}[t]{@{}lrll}
      &\pmb{\gamma}^{(k+1)}
      &=& \argmax{\pmb{\gamma} \geq 0}
        \mathbb{E}_{\mathbf{w}|\mathbf{y}; \pmb{\gamma}^{(k)}, \Sigma^{(k)}} \big[p(\mathbf{y}, \mathbf{w}; \pmb{\gamma}, \Sigma^{(k)}\big] \\
      \Rightarrow
      &\gamma_j^{(k+1)} &=& \frac{1}{LM_j} \sum_{i\in \Lambda_j} \left[ \pmb{\mu}_i^2 + (\Sigma_\mathbf{w})_{i,i} \right], \\[5pt]
      &\Sigma^{(k+1)}
      &=& \argmaxi{\Sigma >0}
        \mathbb{E}_{\mathbf{w}|\mathbf{y}; \pmb{\gamma}^{(k)}, \Sigma^{(k)}} \big[p(\mathbf{y}, \mathbf{w}; \pmb{\gamma}^{(k)}, \Sigma)\big] \\
      \Rightarrow
      &(\sigma_l^2)^{(k+1)}  &=& \frac{1}{N_{\mathbf{y}}} \big\{
             \|\mathbf{y}^{[l]} - \mathbf{A}^{[l]} \pmb{\mu}\|^2 + \\
      &&& (\sigma_l^2)^{(k)} \sum_{i=1}^{M_\mathbf{w}} \big[ 1-\big( \pmb{\gamma}_i^{(k)} \big)^{-1} (\Sigma_\mathbf{w})_{i,i} \big] \big\},
    \end{array}
  \end{equation*}
\end{itemize}
for $j=1,\dots,M$; $i = 1, \dots, M_\mathbf{w}$; $l = 1,\dots,L$
and $\Sigma^{(k+1)} = \diag((\sigma_1^2)^{(k+1)}I_1,\dots,(\sigma_L^2)^{(k+1)}I_L)$,
where $\pmb{\mu}$ and $\Sigma_\mathbf{w}$ are calculated via (\ref{eq:bayes-posterir})
provided with $\pmb{\gamma}^{(k)}$ and $\Sigma^{(k)}$,
$M_j$ is given in (\ref{eq:def-Nk}) and $\Lambda_j$
denotes the row (column) indexes associated with $\gamma_j$ in $\Gamma$.

\subsection{Sparse estimation via sampling}
\label{sec:mcmc-methods-arx}

This section will propose an approach based on sampling methods for sparse
estimation, which is powerful to handle small data and has more flexibility
to be extended. This idea is inspired by the work in Monte Carlo strategies
for binary variables, e.g., \cite{Kuo1998} for model selection, or sampling
methods for Ising and Potts models in statistical physics (e.g., see
\cite{Liu2008}). The sampling methods will be used in this section to
estimate the underlying sparse structures. Note that this approach does not
heuristically enforce sparsity as the $l_1$ methods or SBL does.


Consider
\begin{equation}
  \label{eq:regression-model-sampling}
  \mathbf{y} = \mathbf{A} \mathbf{w} + \pmb{\xi}, \quad \pmb{\xi} \sim \mathcal{N}(0, \Sigma),
\end{equation}
where $\Sigma = \diag(\sigma_1^2I_1, \dots, \sigma_L^2I_L)$ with
$\sigma_l \in \mathbb{R}_+$; identity matrix $I_l$ is of dimension $N_l$
(that denotes the length of time series from the $l$-th experiment);
$\mathbf{y}$ is of dimension $M_\mathbf{y}$; and $\mathbf{w}$ is of dimension
$M_\mathbf{w}$ (it is easy to see $M_\mathbf{y} = \sum_{l=1}^L N_l$ and
$N_\mathbf{w} = \|\rho^{S}\|_1$ from \eqref{eq:regression-form-exp-dataset} and
\eqref{eq:group-sparsity-2types}).
Hence, the likelihood of \eqref{eq:regression-model-sampling} is Gaussian,
given as
\begin{equation}
  \label{eq:likelihood-sampling}
  p(\mathbf{y} \mid \mathbf{w}; \Sigma) =
  \left(2\pi \right)^{-\frac{M_\mathbf{y}}{2}} |\Sigma|^{-\frac{1}{2}}
  \exp
  \left( -\frac{1}{2} \|\mathbf{y}-\mathbf{A}\mathbf{w}\|_{\Sigma^{-1}}^2 \right).
\end{equation}

The key that allows sparse estimation via sampling is to introduce a
``virtual'' variable $\mathbf{s} \triangleq [s_1,\dots,s_M]$ to describe
the underlying sparse structure of $\mathbf{w}$, defined by
\begin{equation}
  \label{eq:w-sG}
  \mathbf{w}
  = \diag(s_1I_1,\dots, s_{M}I_{M}) \pmb{\vartheta}
  \triangleq S \pmb{\vartheta},
\end{equation}
where indicator variables $s_i \in \{0,1\}$, each identity matrix $I_i$
($i=1,\dots,M$) is of dimension specified by $\rho^S$ in
\eqref{eq:block-dim-rho} correspondingly, and
$\pmb{\vartheta} \in \mathbb{R}^{M_\mathbf{w}}$. Now the idea becomes clear
that, as sampling the atomic spins in the Ising model, sampling
$\mathbf{s}$ from any posterior distribution tells the sparsity structure
of $\mathbf{w}$. There could be multiple choices of priors for
$\mathbf{s}$. Here we suggest the Bernoulli distribution,
given as $\mathbf{s} \sim \prod_{i=1}^M \mathcal{B}(1, p_i)$
($p_i \in (0,1)$, e.g., $p_i = 1/2$). And the prior for $\pmb{\vartheta}$
is chosen to be Gaussian,
\begin{equation}
  \label{eq:prior-vartheta}
  p(\pmb{\vartheta}; \Gamma) =
  (2\pi)^{-\frac{M_\mathbf{w}}{2}} |\Gamma|^{-\frac{1}{2}}
  \exp\left(-\frac{1}{2}\pmb{\vartheta}^T \Gamma^{-1} \pmb{\vartheta}\right).
\end{equation}
The covariance structure of $\Gamma$ could either be simply
$\Gamma = \gamma I$ or be diagonal
$\Gamma = \diag(\gamma_1I_1,\dots,\gamma_MI_M)$, in which $I_i$
($i=1,\dots,M$) is the same as \eqref{eq:w-sG}. Letting
$\pmb{\gamma} \triangleq (\gamma_1,\dots,\gamma_M)$, for brevity, we will
use $\Gamma$ and $\pmb{\gamma}$ interchangeably\footnote{If using
  $\Gamma = \gamma I$, $\pmb{\gamma}$ becomes a simple scalar. The
  difference from the SBL is that the importance of $\Gamma$ in sparsity
  pursuit has been largely weaken.}. Furthermore, we use the inverse Gamma
distribution as the hyperprior independently for each $\sigma^2_l$
($l = 1,\dots,L$)\footnote{One may use the scalar $\sigma^2$ if it is fair
  to assume the variances of $e(t)$ in multiple experiments are the same or
  close.} or ${\gamma_i}$ ($i = 1,\dots,M$) (e.g., see \cite{Berger1993}),
denoted by $\sigma^2_l \sim \mathcal{G}^{-1}(a,b)$ and
${\gamma_i} \sim \mathcal{G}^{-1}(c,d)$, where $a,b,c,d$ choose small
values (e.g., $10^{-4}$) to make these hyperprior non-informative.

Since $\pmb{\vartheta}$ is a multivariate continuous variable, sampling
$\pmb{\vartheta}$ could be particularly costly for large dimensions.
Instead, we perform a ``collapse'' operation on the posterior distribution
\cite{VanDyk2015}, and sample $\mathbf{s}$ from the marginalised posterior
to determine the sparse structure. The estimation of $\mathbf{w}$ could be
performed at the end by running a linear regression with its sparsity
structure specified. The posterior
$p(\mathbf{s}, \pmb{\vartheta}, \pmb{\gamma}, \Sigma \mid \mathbf{y})$ is
marginalised over $\pmb{\vartheta}$,
\begin{equation}
  \label{eq:margin-dist-no-theta}
  \begin{array}{l@{\:}l}
    p(\mathbf{s},\pmb{\gamma},\Sigma \mid \mathbf{y})
    &\propto \displaystyle p(\mathbf{s}) p(\pmb{\gamma}) p(\Sigma)
      p(\mathbf{y} \mid \mathbf{s}, \Sigma, \pmb{\gamma}),\\
    &\propto \displaystyle p(\mathbf{s}) p(\pmb{\gamma}) p(\Sigma)
      \int p(\mathbf{y}\mid\mathbf{w},\Sigma)
      p(\pmb{\vartheta}\mid\pmb{\gamma}) \mathrm{d}\pmb{\vartheta}.
  \end{array}
\end{equation}
Substituting the priors and completing squares,
it yields
\begin{equation}
  \label{eq:prob-t-given-s-R-gamma}
  p(\mathbf{y} \mid \mathbf{s}, \Sigma, \pmb{\gamma}) = (2\pi)^{-\frac{M_{\mathbf{y}}}{2}}
  \left| \Sigma_{\mathbf{y}} \right|^{-\frac{1}{2}}
  \exp\left( -\frac{1}{2} \|\mathbf{y}\|_{\Sigma_{\mathbf{y}}^{-1}}^2 \right),
\end{equation}
where
\begin{equation}
  \label{eq:cov-t}
  \Sigma_{\mathbf{y}} = \Sigma + \mathbf{A} S \Gamma S \mathbf{A}^T.
\end{equation}
The onus is to sample $\mathbf{s}, \Sigma$ and $\pmb{\gamma}$ from
$p(\mathbf{s}, \pmb{\gamma}, \Sigma \mid \mathbf{y})$. The
\emph{Metropolis-Hastings} algorithm within the \emph{systematic-scan Gibbs
  sampler} (e.g., see \cite{Liu2008}) will be used to draw samples.
Suppose that the $k$-th samples $\mathbf{s}^{(k)}, \pmb{\gamma}^{(k)}$ and
$\Sigma^{(k)}$ have been available. At the $k+1$ iteration,
\begin{itemize}
\item draw $\mathbf{s}^{(k+1)}$ from the conditional distribution\\
  \begin{math}
    p(\mathbf{s}\mid\mathbf{y}, \pmb{\gamma}^{(k)},\Sigma^{(k)})
  \end{math},
\item draw $\pmb{\gamma}^{(k+1)}$ from $p(\pmb{\gamma}\mid\mathbf{y},
  \mathbf{s}^{(k+1)}, \Sigma^{(k)})$, and
\item draw $\Sigma^{(k+1)}$ from $p(\Sigma \mid\mathbf{y}, \mathbf{s}^{(k+1)}, \pmb{\gamma}^{(k+1)})$.
\end{itemize}
In each sampling step, we use the \emph{Metropolis-Hastings} algorithm.
Consider drawing samples from $p(\mathbf{s}\mid\mathbf{y},\pmb{\gamma},\Sigma)$,
in which the key is to choose a proposal distribution
$T(\mathbf{s}^{(k)},\hat{\mathbf{s}})$, where $\mathbf{s}^{(k)}$ is a given
sample and $\hat{\mathbf{s}}$ is a new
configuration. Here we use a simple scheme that gives a symmetric proposal
distribution (i.e.
$T(\mathbf{s}^{(k)},\hat{\mathbf{s}}) = T(\hat{\mathbf{s}},\mathbf{s}^{(k)})$).
We randomly pick an element of $\mathbf{s}^{(k)}$ and flip it, i.e. draw
$\hat{i}$ from the uniform distribution on $\{1,\dots,M\}$ and set
$\hat{s}_i$ to be equal to $s_i^{(k)}$ if $i \neq \hat{i}$ and $1-s_i^{(k)}$ otherwise.
The Metropolis-Hastings algorithm for
$p(\mathbf{s} \mid \mathbf{y},\pmb{\gamma},\Sigma)$ is given as below: given
$\pmb{\gamma}^{(k)}, \Sigma^{(k)}$ and $\mathbf{s}^{(k)}$,
\begin{itemize}
\item draw $\hat{\mathbf{s}}$ from the proposal distribution
  $T(\mathbf{s}^{(k)}, \hat{\mathbf{s}})$ using the above scheme;
\item draw $U \sim \text{Uniform}[0,1]$ and update
    \begin{equation*}
      \hspace*{-5mm}
      \begin{array}{r@{\:}l}
      &\mathbf{s}^{(k+1)} =
      \begin{cases}
        \hat{\mathbf{s}}, & \text{if } U \leq r(\mathbf{s}^{(k)}, \hat{\mathbf{s}})\\
        \mathbf{s}^{(k)}, & \text{otherwise}
      \end{cases},
      \;\; \text{with} \\[10pt]
      &r(\mathbf{s}^{(k)},\hat{\mathbf{s}}) = \min
      \displaystyle \left\{ 1,
      \frac{p(\hat{\mathbf{s}},\pmb{\gamma}^{(k)},\Sigma^{(k)} \mid \mathbf{y})
        T(\hat{\mathbf{s}},\mathbf{s}^{(k)})}{
        p(\mathbf{s}^{(k)},\pmb{\gamma}^{(k)},\Sigma^{(k)}\mid\mathbf{y})
        T(\mathbf{s}^{(k)},\hat{\mathbf{s}})} \right\}.
      \end{array}
    \end{equation*}
\end{itemize}
In regard to the Metropolis-Hastings algorithms for $\pmb{\gamma}$ and $\Sigma$,
we use random walk sampling
$\hat{\pmb{\gamma}} = \pmb{\gamma}^{(k)} + \mathbf{u}, \hat{\Sigma} =
\Sigma^{(k)} + \diag(v_1I_1, \dots, v_LI_L)$, where $\mathbf{u}$ and ${v_i}$ are
normally distributed with mean zero and fixed variances (which may need to be
tuned to have sound convergence speeds and acceptance ratios), and the
acceptance probabilities are given as, respectively,
\begin{align*}
  r(\pmb{\gamma}^{(k)},\hat{\pmb{\gamma}})
  &= \min \left\{1,
    \frac{p(\mathbf{s}^{(k+1)},\hat{\pmb{\gamma}},\Sigma^{(k)}
    \mid
    \mathbf{y})}{p(\mathbf{s}^{(k+1)},\pmb{\gamma}^{(k)},\Sigma^{(k)}
    \mid\mathbf{y})}
    \right\}, \\
  r({\Sigma}^{(k)},\hat{\Sigma})
  &= \min \left\{1,
    \frac{p(\mathbf{s}^{(k+1)},\pmb{\gamma}^{(k+1)},\hat{\Sigma}
    \mid\mathbf{y})}{p(\mathbf{s}^{(k+1)},\pmb{\gamma}^{(k+1)},\Sigma^{(k)}
    \mid\mathbf{y})}
    \right\}.
\end{align*}

\section{Numerical examples}
\label{sec:numerical-examples}


This section presents several Monte Carlo studies to show inference
performance from multiple experiments. It first applies three sparsity
techniques in the proposed method for multiple experiments, presented in
Section~3 \& 4.
To further show the superiority of simultaneous processing of multiple-experiment
data, we provide two more studies that compare the proposed method with the
``naive'' treatment (i.e. inferring networks separately from each dataset,
and then choosing the edges in common) widely used in applications.

\emph{ARX networks} refer to such a class of discrete-time DSF models
\eqref{eq:dynet-model-description} that each MISO transfer function
\eqref{eq:dynet-model-description-ith} can be exactly rewritten as an ARX
model. To randomly generate proper signals for benchmarking, it is further
required that these random ARX networks should be \emph{stable} (see
\cite{Yue2018} or \cite{Yue2019}; or only demanding BIBO stability) and
have sparse network structures that should not be oversimplified (e.g.,
trees or acyclic digraphs may not be satisfactory). This is particularly
challenging due to the demands on both sparse structures (with loops) and
stability.
Here we propose a simplified strategy for random model generation, whose
details are provided in the supplement
\cite[app.~\ref{appdix:arx-generation}]{Yue2016a} for interested readers.
The basic idea is to repetitively apply the \emph{small gain theorem} in
sequence so as to stabilise the whole ARX network. Moreover, to generate
its replica models for multiple experiments (i.e. models with $L > 1$),
we randomly perturb its nonzero model parameters and/or noise
covariance. Due to technical issues in model generation, the perturbation
remains relatively minor to avoid triggering network instability.

The performance indexes for benchmarks use the \emph{precision} (Prec) and
the \emph{true positive rate} (TPR), defined as
$\text{Prec} = {\text{TP}}/{(\text{TP}+\text{FP})}$ and
$\text{TPR} = {\text{TP}}/{(\text{TP}+\text{FN})}$, where TP (true
positive), FP (false positive) and FN (false negative) are the standard
concepts in the \emph{Receiver Operating Characteristic} (ROC) curve or the
\emph{Precision Recall} curve (e.g. see \cite{Sahiner2017}). One may
understand Prec as the percentage of correct arcs in the inferred network,
and TPR as the percentage of correctly inferred arcs in the ground truth.
The ideal is to achieve both high values, while, if not possible, Prec has
higher priority since its low value results in completely useless
inference.

The benchmark study, as shown in Fig.~\ref{fig:arxnet-perf} (and
Table~\ref{tbl:arx-dynet} in the supplement \cite{Yue2016a}), performs the
proposed reconstruction method for randomly generated data, equipped with
iterative reweighted $l_1$ method (labeled as ``GIRL1''), sparse Bayesian
learning (labeled as ``GSBL'') and the sampling method (labeled as ``GSMC''
) for group sparsity. There are two test scenarios: 1) models with $p = 10$
(i.e. $\#\text{nodes} = 10$) and different SNRs
($\text{SNR} = 0, 10, 20, 40$ dB); 2) models with $\text{SNR} = 10$ dB and
different number of nodes $p = 5, 10, 15, 20$. Each test (with given $p$
and $\text{SNR}$) generates $50$ random stable ARX networks and simulates
time series of length $1000$. All models are set to the same sparsity
density $0.2$, i.e. the total number of nonzero entries in $Q$ is $0.2p^2$.
The input signals are independently and identically distributed Gaussian
noise with zero mean and unit variance. In the test of GIRL1, the
regularization parameter $\lambda$ is set to as follows:
$\lambda = 0.1, 0.1, 0.01, 0.001$ for
$\text{SNR} = 0, 10, 20, 40 \text{dB}$, respectively; and
$\lambda = 0.05, 0.1, 0.1, 0.1$ for $p = 5, 10, 15, 20$. The parameter
$\lambda$ is chosen roughly among values in logarithmic scales for one
model by reviewing the sparsity density of the inference results (assuming
we roughly know how sparse the network should be), and then is used for all
50 models. One certainly can apply cross-validation or bootstrap methods to
choose and tailor $\lambda$ for every model for better performance.

The comparative results in Fig.~\ref{fig:arxnet-perf} tell that GSMC
overall outperform in terms of balance between Prec and TPR. And the users
are free from tuning regularisation parameters as required in GIRL1 or
other regularisation methods. Moreover, GSMC turns to be impressively
powerful when time series of limited lengths are available, whereas GIRL1
and GSBL could fail completely. However, the computational cost might
increase dramatically when the problem scale increases. We strongly suggest
GSMC when problem scales are not too large or the length of time series is
particularly limited. The superiority of GIRL1 (or group LASSO) could be
the freedom of tuning regularisation parameters, which allows us to
heuristically handle large model certainties or non-Gaussian noises in
practice. And when the Prec and TPR cannot be both satisfactory, a
conservative choice of $\lambda$ helps to improve Prec at the expense of
TPR in order to ensure inferred edges being reliable. This explains, in
Fig.~\ref{fig:arxnet-perf}, GIRL1 has better performance on Prec (larger
means and smaller variances) than GSBL over different SNRs or \#nodes;
while the performance on TPR shows slighter weaker than GSBL. This
comparison for GSBL shows a slight increase of Prec or TPR when the SNR
increases, which, however, is not as significant as our intuition might
tell. The reason is that both GSBL and GSMC in theory can handle data with
a reasonably large range of SNRs by estimating noise variances reliably.



\begin{figure}[htbp]
  \centering

  \begin{subfigure}[b]{0.45\textwidth}
    \centering
    \includegraphics[width=\textwidth]{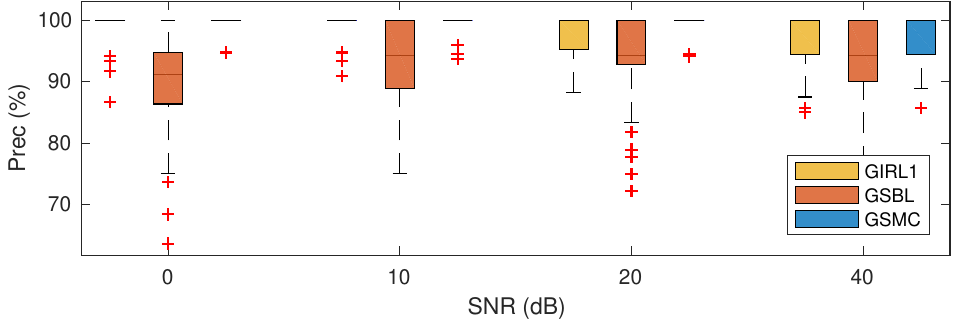}
  \end{subfigure}
  \\[5pt]
  \begin{subfigure}[b]{0.45\textwidth}
    \centering
    \includegraphics[width=\textwidth]{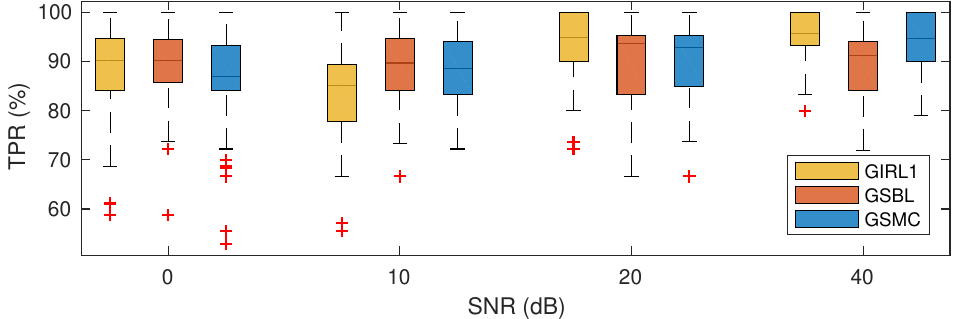}
    \caption{multiple SNRs}
    \label{subfig:boxplot-mSNRs}
  \end{subfigure}
  \\
  \begin{subfigure}[b]{0.45\textwidth}
    \centering
    \includegraphics[width=\textwidth]{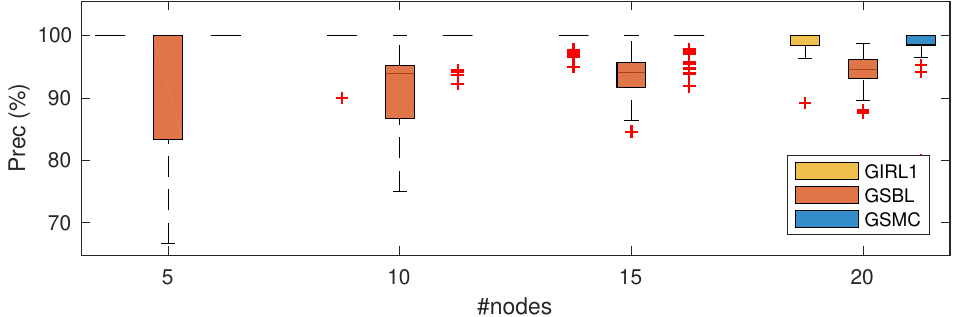}
  \end{subfigure}
  \\[5pt]
  \begin{subfigure}[b]{0.45\textwidth}
    \centering
    \includegraphics[width=\textwidth]{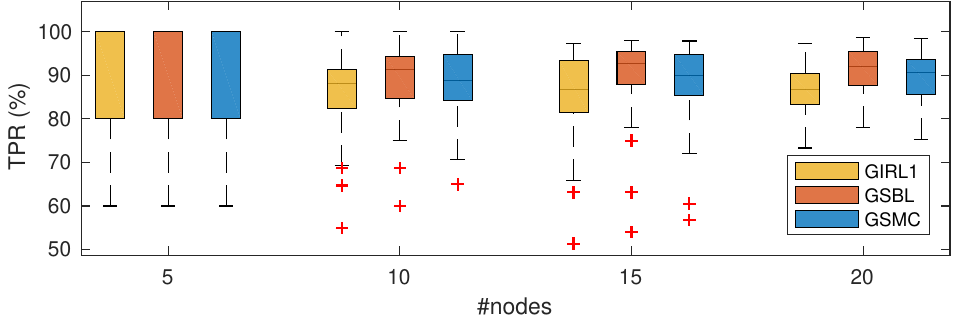}
    \caption{multiple \#nodes}
    \label{subfig:boxplot-mNodes}
  \end{subfigure}

  \caption{Comparative performance of network reconstruction using GIRL1,
    GSBL and GSMC in scenarios with different SNRs or \#nodes.}
  \label{fig:arxnet-perf}
\end{figure}

\begin{figure}[htbp]   
  \centering
  \includegraphics[width=.4\textwidth]{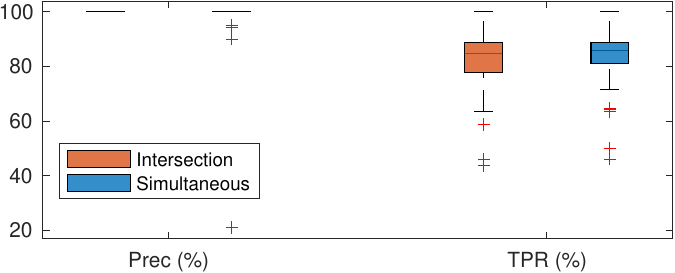}
  \caption{Comparative study of the proposed method (labelled
    ``Simultaneous'') and the ``naive'' treatment (``Intersection'') to
    heterogeneous datasets with $L = 2$.}
  \label{fig:boxplot-intersection}
\end{figure}

\begin{figure}[htbp]   
  \centering
  \includegraphics[width=.4\textwidth]{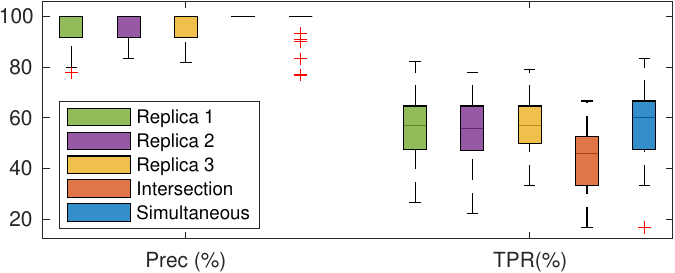}
  \caption{Comparative study to show the superiority of the proposed method
    to the separate processing of datasets, where $L = 3$ and the length of
    time series is restricted to $50$.}
  \label{fig:boxplot-separate-treat}
\end{figure}

To further show the superiority of the treatment of heterogeneous data
presented in Section~\ref{sec:heterogen-data}, we perform two more Monte
Carlo studies, in which GSMC is adopted for sparsity. One sets
$\text{SNR} = 0 \text{ dB}$, $L = 2$ and prepares time series of length
$1000$ for inference. The comparative result in
Fig.~\ref{fig:boxplot-intersection} applies two ways to infer networks for
multiple experiments, where one use the proposed method, and the other
infers $2$ networks separately from each replica data and then takes
intersection of their edge sets. The result shown in
Fig.~\ref{fig:boxplot-intersection} is easy to understand in theory. The
intersection gains confidence on the inferred edges, i.e. higher Prec,
while sacrificing TPR (its TPR is always equal to or smaller than any TPR
of inference from a single replica). In practice, if the diversity between
replica data is significant or the number of replica is large, this
``naive'' treatment becomes very inefficient in the sense that it may miss
a large amount of edges that could be explored from data. This is due to
approximate methods for sparsity, and thus the heterogeneity of datasets
could lead to different topology if processing datasets separately. Another
Monte Carlo study manifests that the proposed method could improve both
Prec and TPR, as shown in Fig.~\ref{fig:boxplot-separate-treat}, when
single dataset is deficient. It restricts the length of time series to $50$
and sets $L =3$. Fig.~\ref{fig:boxplot-separate-treat} shows, by processing
three replica data together in the proposed way, we manage to push Prec
close to $100\%$ while not affecting TPR or even slightly improving TPR.

\section{Conclusions}
\label{sec:conclusions}

This paper discusses dynamic network reconstruction from heterogeneous
datasets in the framework of dynamical structure functions (DSFs). It has
been addressed that dynamic network reconstruction for linear systems can
be formulated as identification of DSFs with sparse structures. To take
advantage of heterogeneous datasets from multiple experiments, the proposed
method integrates all datasets in one regression form and resorts to group
sparsity to guarantee network topology being consistent over replica. To
solve the optimisation problem, the treatments using classical convex
approximation, SBL and sampling methods have been introduced and extended.
The numerical examples manifest the performance of reconstruction from
heterogeneous data and reveal practical experience that could guide
applications.

\begin{ack}
  This work was supported by Fonds National de la Recherche Luxembourg
  (Ref.~AFR-9247977 and Ref.~C14/BM/8231540), and partly supported by the 111 Project on Computational
  Intelligence and Intelligent Control under Grant B18024.
\end{ack}

\bibliographystyle{unsrt}
\bibliography{./ref/library}           

\appendix


\section{Dynamical structure functions}
\label{appdix:dsf}

Consider a dynamical system given by the discrete-time state-space representation in the innovations form
\begin{equation}
  \label{eq:ss-sys}
  \begin{array}{l@{\;}l}
    x(t_{k+1}) &= A x(t_k) + B u(t_k) + K e(t_k),\\
    y(t_{k})   &= C x(t_k) + D u(t_k) +   e(t_k),
  \end{array}
\end{equation}
where $x(t_k)$ and $y(t_k)$ are real-valued $n$ and $p$-dimensional random
variables, respectively; $u(t_k) \in \mathbb{R}^m$, $A, B, C, D, K$ are of
appropriate dimensions; and $\{e(t_k)\}_{k \in \mathbb{N}}$ is a sequence
of i.i.d. $p$-dimensional random variables with
$e(t_k)\sim \mathcal{N}(0,R)$. The initial state $x(t_0)$ is assumed to be
a Gaussian random variable with unknown mean $m_0$ and variance $R_0$.
Without loss of generality, we assume $n \geq p$ and $C$ is of full row
rank.  The procedure to define the DSFs from \eqref{eq:ss-sys} mainly
refers to \cite{Goncalves2008,Chetty2015}. Without loss of generality,
suppose that $C$ is full row rank (see \cite{Chetty2015} for a general
$C$). Create the $n\!\times\!n$ state transformation $T = \bm{C^T & E}^T$,
where $E \in \mathbb{R}^{n\times (n-p)}$ is any basis of the null space of
$C$ with $T^{-1} = \bm{\bar{E} & E}$ and $\bar{E} = C^T(C C^T)^{-1}$. Now
we change the basis such that $z = Tx$, yielding $\hat{A} = TAT^{-1}$,
$\hat{B} = TB$, $\hat{C} = CT^{-1}$, $\hat{D} = D$, $K = TK$, and
partitioned commensurate with the block partitioning of $T$ and $T^{-1}$ to
give
\begin{equation*}
  \begin{array}{r@{\:}c@{\:}l}
    \bm{z_1(t_{k+1}) \\ z_2(t_{k+1})} &=&
    \begin{bmatrix}
      \hat{A}_{11} & \hat{A}_{12} \\ \hat{A}_{21} & \hat{A}_{22}
    \end{bmatrix}
    \bm{z_1(t_k) \\ z_2(t_k)} +
    \bm{\hat{B}_1 \\ \hat{B}_2} u(t_k)
    + \bm{\hat{K}_1 \\ \hat{K}_2} e(t_k), \\
    y(t_k) &=& \bm{I & 0} \bm{z_1(t_k) \\ z_2(t_k)} + D u(t_k) + e(t_k).
  \end{array}
\end{equation*}
Introduce the shift operator $q$ and solve for $z_2$,
yielding $qz_1(t_k) = W(q) z_1(t_k) + V(q) u(t_k) + L(q) e(t_k) $, where
$W(q) = \hat{A}_{11} + \hat{A}_{12}(qI - \hat{A}_{22})^{-1} \hat{A}_{21}$,
$V(q) = \hat{B}_{1} + \hat{A}_{12}(qI - \hat{A}_{22})^{-1} \hat{B}_{2}$, and
$L(q) = \hat{H}_{1} + \hat{A}_{12}(qI - \hat{A}_{22})^{-1} \hat{H}_{2}$.  Let
$D_{W}(q) = \diag(W(q))$ be a diagonal matrix function composed of the diagonal
entries of $W(q)$. Define $\hat{Q}(q) = (qI - D_W)^{-1}(W-D_W)$,
$\hat{P}(q) = (qI - D_W)^{-1}V$, and $\hat{H}(q) = (qI - D_W)^{-1}L$, yielding
$z_1(t_k) = \hat{Q}(q) z_1(t_k) + \hat{P}(q) u(t_k) + \hat{H}(q) e(t_k)$.
Noting that $z_1(t_k) = y(t_k) - Du(t_k) - e(t_k)$, the DSF of \eqref{eq:ss-sys}
with respect to $y$ is then given by
\begin{equation}
  \label{eq:qph-general-dsf}
  \begin{array}{l@{\:}l}
    Q(q) &= \hat{Q}(q), \\
    P(q) &= \hat{P}(q) + (I-\hat{Q}(q))D,\\
    H(q) &= \hat{H}(q) + (I - \hat{Q}(q)).
  \end{array}
\end{equation}
Noting that the elements of $\hat{Q}, \hat{P}, \hat{H}$ (except zeros in the
diagonal of $\hat{Q}$) are all strictly proper, it is easy to see that $Q$ is
strictly proper and $P,H$ are proper. It has been proven in \cite{Chetty2015}
that the DSF defined by this procedure is invariant to the class of block
diagonal transformations used above, which implies it is a feasible extension of
the definition of DSFs given in \cite{Goncalves2008} for the particular class of
state-space models with $C = \bm{I & 0}, D = 0$.

\section{Indexing in parametrization}
\label{appdix:addit-deta-param}

To implement the (extended) group LASSO, one may need to find all elements of
$\mathbf{w}$ that correspond to the $ k^{E}$-th small group of $\mathbf{w}$
(i.e. the $k^{E}$-th vector $\mathbf{w}_k^{[l]}$ in $\mathbf{w}$) or the
$k^{S}$-th large group of $\mathbf{w}$ (i.e. the vector $\mathbf{w}_{k^S}$).
Here are the formulas:
\begin{itemize} 
\item
  \(
  k = (C \sum_{j=1}^{\lceil k^{E}/C \rceil -1} \rho_j) \cdot \mathbf{1} +
  \big[ \big((k^{E}-1)\!\!\mod\!C\big) \rho_{\lceil k^{E}/C \rceil} +1  \big]
  \!:\!1\!:\!
  \big((k^{E}-1)\!\!\mod\!C + 1\big) \rho_{\lceil k^{E}/C \rceil}
  \)
\item
  \(
  k = \Big( C \sum_{j=1}^{k^{S}-1} \rho_j\Big) \cdot \mathbf{1} +
  1\!:\!1\!:\!C \rho_{k^{S}}
  \)
\end{itemize}
where $\rho$ is given in \eqref{eq:block-dim-rho}, $\mathbf{1}$ is a row
vector of 1's of the matched dimension, $m\!:\!1\!:\!n\ (m, n \in
\mathbb{N}_+)$ denotes a row vector $[m, m+1, \dots, n]$, and $\lceil x
\rceil$ denotes the smallest natural number that is larger than $x$.


\section{Random generation of ARX networks}
\label{appdix:arx-generation}

Model generation of ARX networks is not straightforward due to the requirements
of stability and sparsity. First we need to clarify these three words that
quantifies or specifies the ARX networks our study: ``random'', ``sparse'' and
``stable''.  The word ``random'' demands that both network structures and model
parameters are randomly generated. The sparsity demands the network structures
of ARX models being sparse; meanwhile we avoid such sparse structures that the
network degenerates into a family of separate small networks, or the network has
oversimplified structures, e.g. a tree with depth 2 (models generated by
\texttt{drmodel.m} in MATLAB).  Our simulation ensures that the generated
networks are not acyclic, since the feedback loops
are essential in physical systems but challenging to deal with in network
reconstruction. The stability of ARX networks derives from the definition of
network stability for DSFs (see \cite[chap.~2.3]{Yue2018}), which requests that
each polynomial in $A(q)$ (our simulation also includes $B^y(q)$ and $B^u(q)$)
is stable (i.e. all roots stay inside of the unit circle on the complex plane),
and the resultant MIMO transfer function (from $u$ to $y$) is stable. The
difficulty on model generation is due to the latter requirement, since the DSF
model with random sparse network topology may not be BIBO stable even if each
entry (SISO transfer function) in $Q$ and $P$ has been chosen to be stable and
even minimal-phase.

\begin{figure}[htbp]   
  \centering
  \includegraphics[width=.45\textwidth]{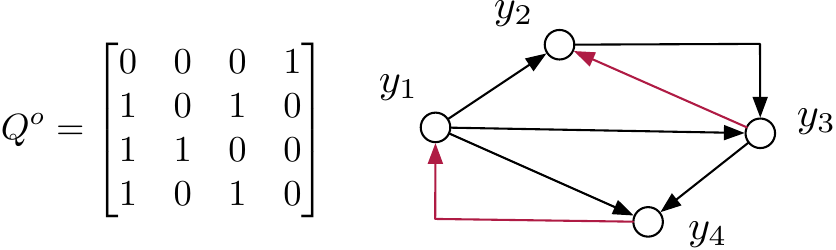}
  \caption{An example of the Boolean matrix and its digraph.}
  \label{fig:arxnet-example}
\end{figure}

\begin{figure}[htbp]   
  \centering
  \includegraphics[width=.5\textwidth]{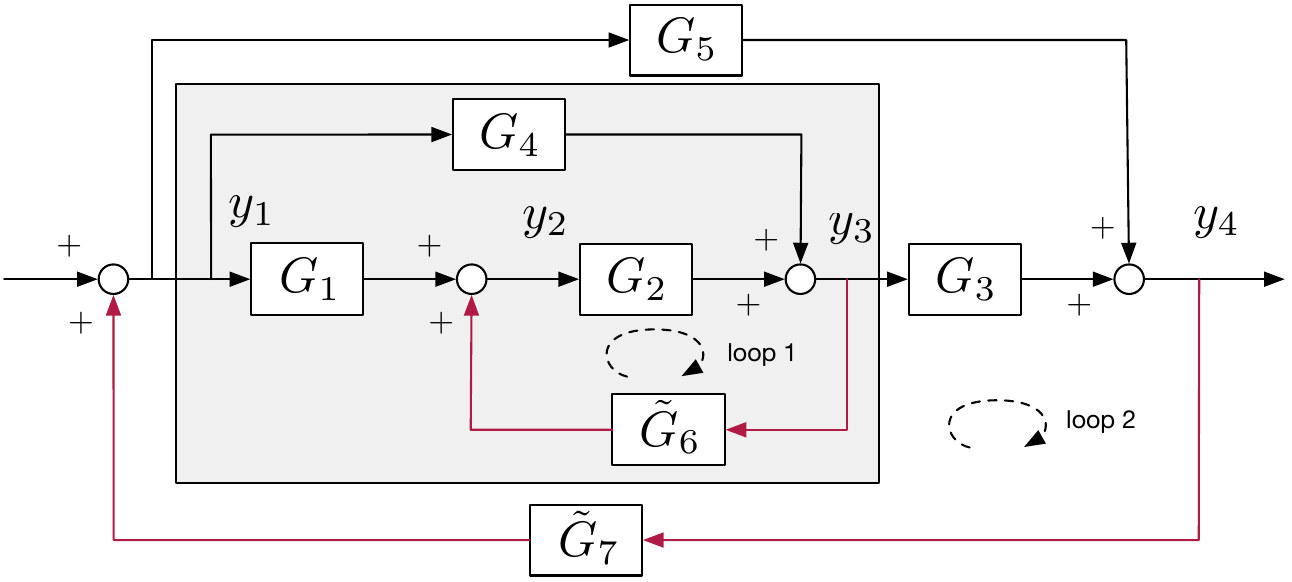}
  \caption{Block diagram of example \eqref{eq:eg-tf-Q} (neglecting inputs).}
  \label{fig:arxnet-bg-1}
\end{figure}

\begin{figure}[htbp]
  \centering

  \begin{subfigure}[b]{0.28\textwidth}
    \centering
    \includegraphics[width=\textwidth]{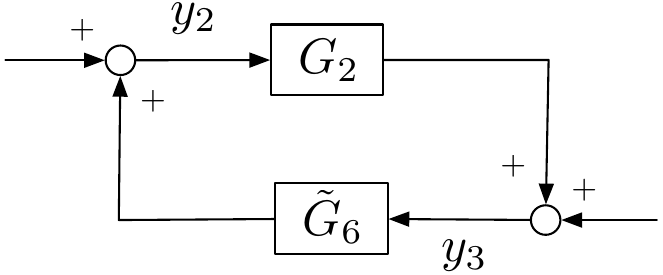}
    \caption{lump-sum system}
    \label{subfig:arxnet-bg-sg-1}
  \end{subfigure}
  \\
  \begin{subfigure}[b]{0.3\textwidth}
    \centering
    \includegraphics[width=\textwidth]{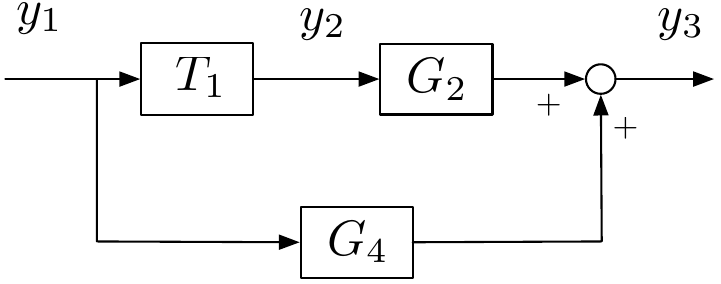}
    \caption{intermediate subsystem}
    \label{subfig:arxnet-bg-interm}
  \end{subfigure}
  \\
  \begin{subfigure}[b]{0.5\textwidth}
    \centering
    \includegraphics[width=\textwidth]{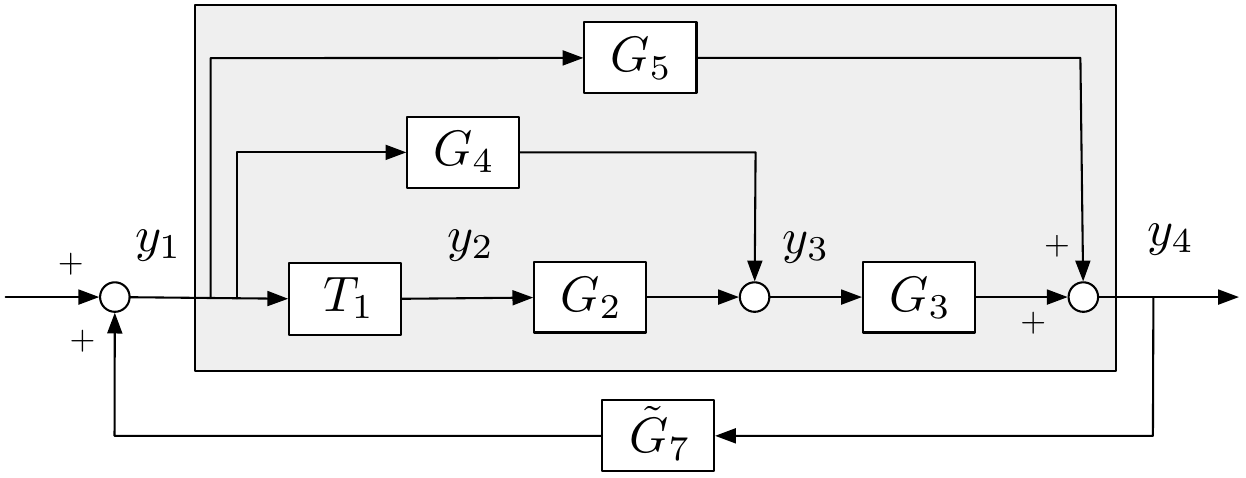}
    \caption{intermediate system}
    \label{subfig:arxnet-bg-2}
  \end{subfigure}
  \\
  \begin{subfigure}[b]{0.28\textwidth}
    \centering
    \includegraphics[width=\textwidth]{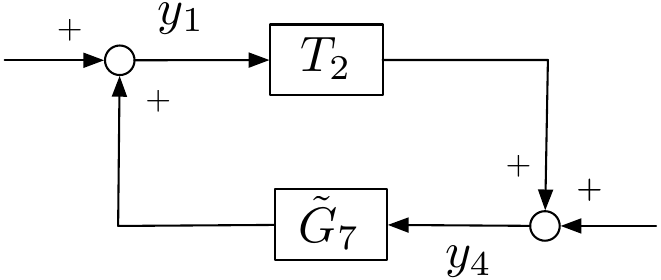}
    \caption{lump-sum system}
    \label{subfig:arxnet-bg-sg-2}
  \end{subfigure}

  \caption{List of intermediate systems that assist the description of the whole
  procedure of stabilizing ARX networks in model generation.}
  \label{fig:arxnet-intermediates}
\end{figure}

The idea to guarantee stable ARX networks is to repetitively apply the
\emph{small gain theorem} (e.g., see \cite[p.~137]{Zhou1998}). First we generate
a random sparse Boolean matrix $Q^o$, which specifies the network structure (to
ease later discussions, we use the transpose of adjacency matrices), with zero
diagonal and nonzero values of $Q^o_{k,k-1}$, e.g., Fig.~\ref{fig:arxnet-example}.
The lower triangular part of $Q^o$ specifies a block diagram (with unknown
transfer functions) with only feedforward paths\footnote{By default, we use the
  order $y_1 \rightarrow y_2 \rightarrow y_3 \rightarrow y_4$ as the forward
  path. It is certainly free to define any order as the default forward path.},
and the upper triangular part adds feedback loops (marked in red in
Fig.~\ref{fig:arxnet-example}). The key is to use the \emph{small gain
  theorem} to tune the added feedback transfer functions one by one to guarantee
BIBO stability. The transfer matrix $Q(q)$ with structure $Q^o$ is created by
generating random stable polynomials for nonzero entries in $A, B^y$ and $B^u$,
where nonzero entries of $B^y$ are specified by $Q^o$, and then using
\eqref{eq:dynet-vs-armax}.
The last step is to tune the gain of each feedback transfer function, as
specified by $Q^o_U$, using the \emph{small gain theorem} sequentially. Here we
will present an example to show the whole procedure. Suppose that the random
structure for the 4-node network is specified by $Q^o$ and
$P^o = [1\ 0 \ 0 \ 0]^T$ (the Boolean matrix of $P$), and the block diagram is
shown in Fig.~\ref{fig:arxnet-bg-1}, where SISO transfer functions
($G_1(q), \dots, G_8(q) \in \mathcal{RH}_{\infty}$) are generated as
aforementioned,
\begin{equation}
  \label{eq:eg-tf-Q}
  Q =
  \begin{bmatrix}
    0 & 0 & 0 & \beta G_7(q) \\
    G_1(q) & 0 & \alpha G_6(q) & 0 \\
    G_4(q) & G_2(q) & 0 & 0 \\
    G_5(q) & 0 & G_3(q) & 0
  \end{bmatrix},
  P =
  \begin{bmatrix}
    G_8(q) \\ 0 \\0 \\ 0
  \end{bmatrix},
\end{equation}
and $\alpha, \beta$ are positive real numbers that need to be tuned to
guarantee stability, $\tilde{G}_6 \triangleq \alpha G_6$,
$\tilde{G}_7 \triangleq \beta G_7$.
We start with the inner loop (labeled as ``loop 1'' in
Fig.~\ref{fig:arxnet-bg-1}) that is formed by the feedback item
$\tilde{G}_6$, and applied the \emph{small gain theorem} to the interconnected
system shown in Fig.~\ref{subfig:arxnet-bg-sg-1}, which tells to choose
$\alpha < 1/(\|G_2\|_{\infty} \|G_6\|_{\infty})$. We then redraw
Fig.~\ref{fig:arxnet-bg-1} to remove the feedback path of $\tilde{G}_6$, as
shown in Fig.~\ref{subfig:arxnet-bg-interm}, and the resultant whole block
diagram turns to be Fig.~\ref{subfig:arxnet-bg-2}, where
$T_1 = \frac{G_1+ G_4G_6}{1 - G_2\tilde{G}_6}$. Next we compute the lump-sum
transfer function of all feedforward paths from $y_1$ to $y_4$ and present the
following interconnected system in Fig.~\ref{subfig:arxnet-bg-sg-2}, where
$T_2 = (T_1G_2 + G_4)G_3 + G_5$. By applying the \emph{small gain theorem}, we
choose $\beta < 1/(\|T_2\|_{\infty} \|G_7\|_{\infty})$. As demonstrated by this
example, the idea is to tune the feedback component sequentially, from inner
loops (e.g., ``loop 1'' in Fig.~\ref{fig:arxnet-bg-1}) to outer loops (e.g.,
``loop 2''), using the \emph{small gain theorem} to guarantee the internal
stability.

This example uses the block diagrams to explain the procedure to apply the
\emph{small gain theorem}. In practice, the signal-flow graph is a better choice
to represent complicated interconnections and helps to apply \emph{Mason's gain
  formula} to automate the computation of interconnected transfer functions. In
our simulation, to ease the computation, up to networks of 20 nodes (see
experiments for Fig.~\ref{subfig:boxplot-mNodes}), we add at most 3 feedback
paths (since the path/loop finding is an NP problem and cost considerable
time). Moreover, to simply computation, these feedback loops are either
non-touching (no common nodes) or one is ``contained'' by the other (i.e., all
the nodes in loop 1 appear in loop 2). Due to page limits, the general
algorithms will be present in another paper to handle arbitrary feedback and
feasibility of network stabilization.  Before moving to the inference part,
there is one implementation detail in MATLAB deserving to be shared. When you
use \emph{control system toolbox} in MATLAB and deal with many (e.g., $p >=10$)
interconnections (connect in series or parallel, feedback), it may not be a good
idea to compute the lump-sum transfer functions first, e.g., $T_2$ in
Fig.~\ref{subfig:arxnet-bg-sg-2}, and then compute their infinity norms to
apply the \emph{small gain theorem}. The reason is that, even if at each step
you guarantee the subsystems are correctly stabilized, the numeric errors in
model interconnection using \emph{control system toolbox} may lead to the
resultant system being unstable (in theory it should be stable) and you are no
longer able to continuing applying the \emph{small gain theorem}. The solution
used in our simulation is that, instead of computing the lump-sum transfer
function, we compute the infinity norm of each transfer function (e.g.,
$T_1, G_2, G_3, G_4, G_5$ for $T_2$) first and then use \emph{Mason's gain
  formula} to compute an upper bound of $\|T_2\|_{\infty}$.

\section{Random generation of  DSFs and benchmark}
\label{subsec:sim-dynam-struct-funct}

This section considers a Monte Carlo study of 50 runs of inference of
\emph{random stable sparse} networks (DSFs) with 40 nodes. In regard to the
adjectives for DSFs, here are further explanations:
\begin{itemize}
\item \emph{random}: the DSF model in each run is randomly chosen (both network
  topology and model parameters);
\item \emph{stable}: the DSF model is stable, i.e. all poles of $Q, P$ and $H$
  have negative real parts, and all transmission zeros of $(I-Q)$ have negative
  real parts; (see \cite[chap.~2.3]{Yue2018})
\item \emph{sparse}: the number of arcs of the network is much less than that of
  a complete digraph.
\end{itemize}
The numerical example emulates the applications in practice, where the
underlying systems evolves continuously in time and the proposed method uses
parametric approaches to estimate network structures. Hence, we simulate the
continuous-time DSF models and then sample the simulated signals with a chosen
sampling frequency to acquire measurements for later network reconstruction.
More details on the procedure is presented as below:
\begin{enumerate}[label=\arabic*) ]
\item The DSF model will be simulated via its state space realization (both in
  continuous time).  We randomly generate highly sparse stable $A$ matrices (of
  dimension $80\!\times\!80$) for state space models, and
  $B = [0, \dots,1,\dots, 0]^T$, $C = [I_{40 \times 40}\; 0]$,
  $D = 0$. The systems in replica are obtained by perturbing nonzero entries in
  the $A$ matrix.

\item The ground truth networks are calculated by the definition of DSF
  using $(A, B, C, D)$ (see \cite{Goncalves2008}).

\item A step signal is chosen to be the input\footnote{Here we choose to have
    only one input and use a step signal to simulate the biological data. In
    biological experiments, we usually do not have many controlled inputs, and
    most of them are simple signals, like fixed temperatures, adding/removing
    light, fixed pH values, etc.}, and each state variable is perturbed by a
  Gaussian i.i.d. (i.e. process noises). The replica data is acquired by
  randomly perturbing non-zero elements in $A$ and performing simulation. The
  stochastic differential equation is numerically solved by using \texttt{sim.m}
  (choosing the Euler-Maruyama method) from \emph{system identification toolbox} in
  MATLAB. The sampling frequency is chosen to be 40 times of the
  critical frequency of system aliasing (see \cite{Yue2016b}).
\end{enumerate}
The setup of DSF models makes network inference particularly challenging.  There
may exist many loops due to feedback, whose sizes are fairly random. Moreover,
we fill nonzero values in the position $A_{i, i+1}$ of the $A$-matrix to ensure
each network will not degenerate into a family of separate small ones.

\begin{figure}[htb]
  \centering

  \begin{subfigure}[b]{0.23\textwidth}
    \includegraphics[width=\textwidth]{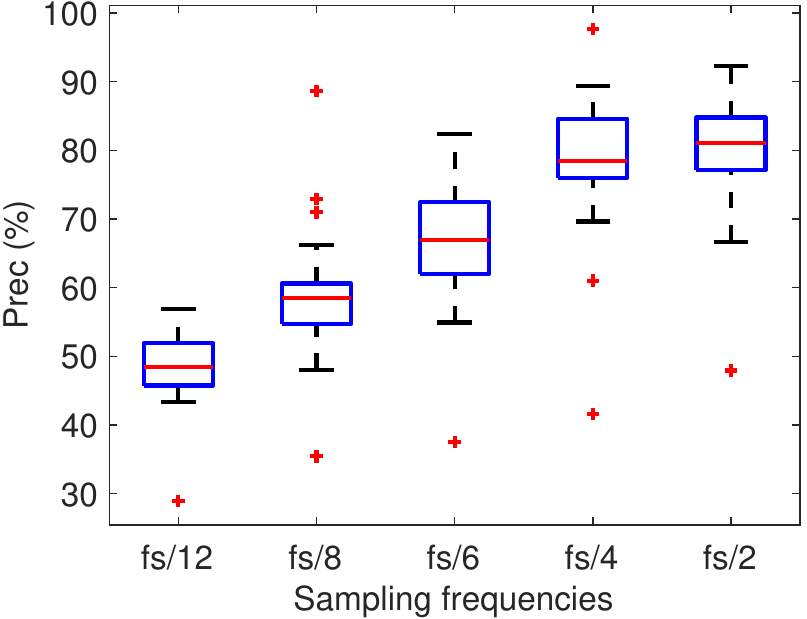} 
  \end{subfigure}
  ~
  \begin{subfigure}[b]{0.23\textwidth}
    \includegraphics[width=\textwidth]{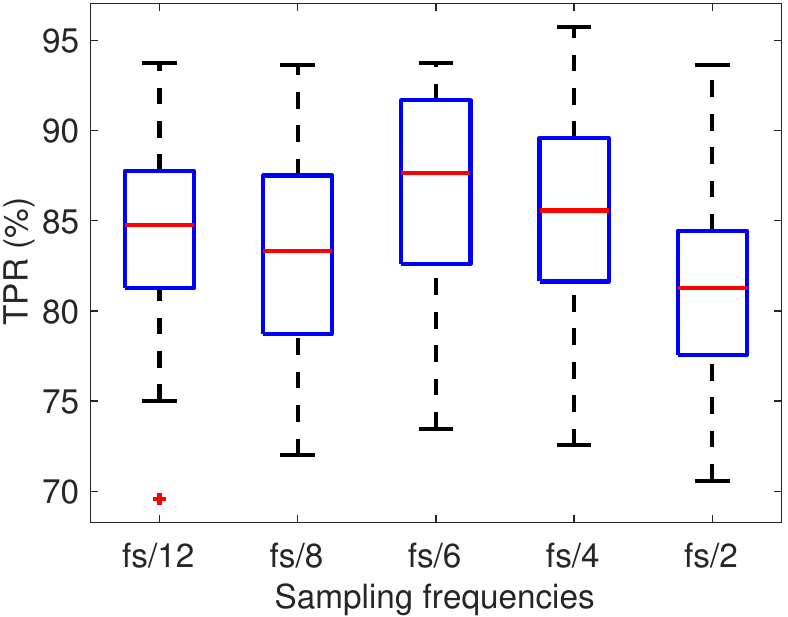} 
  \end{subfigure}
  \\[2ex]
  \begin{subfigure}[b]{0.23\textwidth}
    \includegraphics[width=\textwidth]{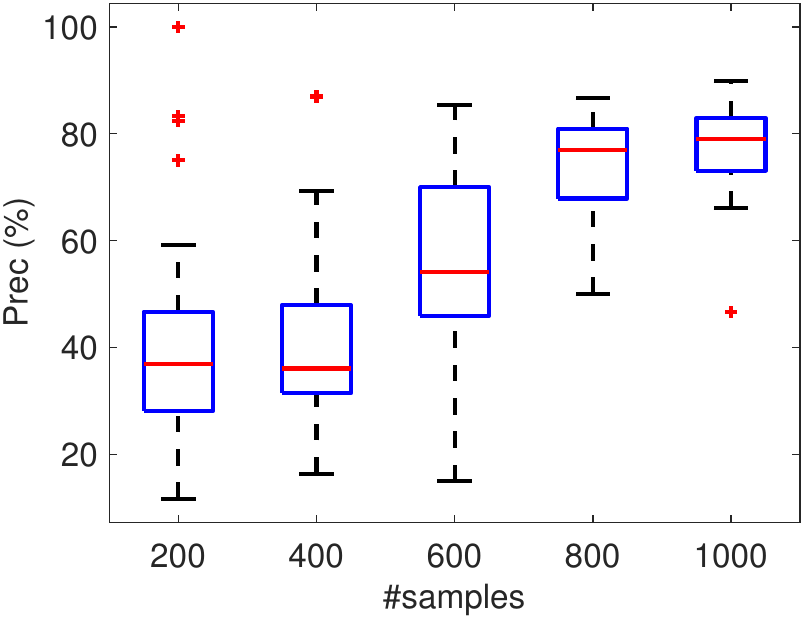} 
  \end{subfigure}
  ~
  \begin{subfigure}[b]{0.23\textwidth}
    \includegraphics[width=\textwidth]{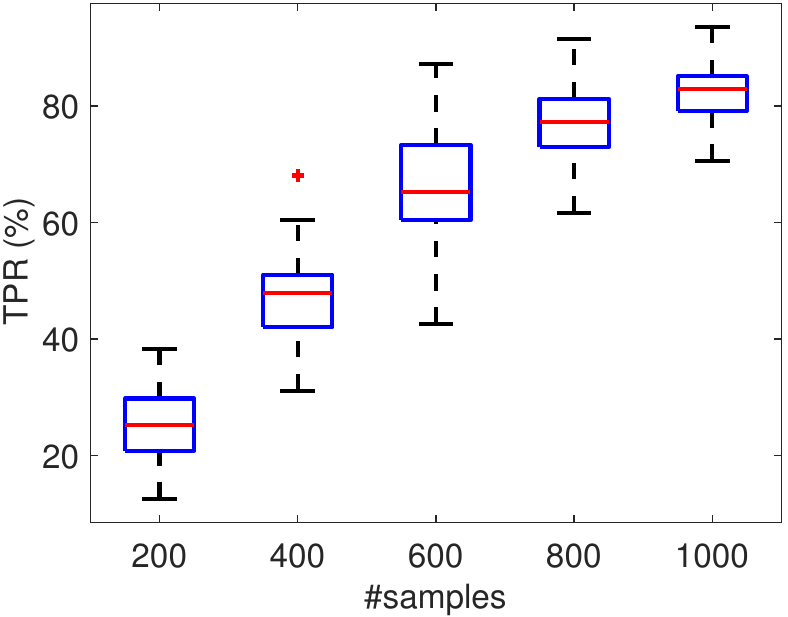} 
  \end{subfigure}

  \caption{Performance of the proposed method on 50 random networks.  The value
    of $\lambda$ is chosen by performing cross-validation on one network.  The
    sampling frequency $f_s$ is the base value used in system simulations. The
    data used in reconstruction are sampled from the simulated signals.  Here
    we use 2 replica datasets, e.g. ``\#samples = 800'' implies each dataset has
    400 samples.
  }
  \label{fig:pred-performance}
\end{figure}

We apply the iterative reweighted $l_1$ method for group sparsity. The SBL and
the sampling method fail mostly in the benchmark of random DSFs due to large
modelling uncertainty using ARX parametrization.  There is a ``trade-off''
between Prec and TPR when selecting regularization parameters $\lambda$. In
theory, there could be an optimal value of $\lambda$ that gives large values of
both Prec and TPR. However, in practice, the Prec is more critical in the sense
that it has to be large enough to keep results useful. Otherwise, even if the
TPR is large, the result will predicate too many wrong arcs to be useful in
applications. As a rule implied from Fig.~\ref{fig:pred-performance}, in
practice, we may choose a conservative value of $\lambda$ to make sure that we
could have most predictions of arcs correctly; then, if more links need to be
explored, we could decrease $\lambda$ to get more connections covered.

As known in biological data analysis, time series are usually of low sampling
frequencies and have limited numbers of samples. To address the importance of
these factors, we run the proposed method over a range of values, shown in
Fig.~\ref{fig:pred-performance}. The sampling frequency is critical for applying
discrete-time approaches for network inference (see \cite{Yue2016f}), which can
be shown by Fig.~\ref{fig:pred-performance}.  The simulation also tells that
the length of time series that is at least four times larger than the number of
unknown parameters could be a fair choice in practice for network
reconstruction.

\section{Supplementary benchmark results}
\label{appdix:benchmark-results}

The detailed result of the benchmark Fig.~\ref{fig:arxnet-perf} in
Section~\ref{sec:numerical-examples} is presented in
Table~\ref{tbl:arx-dynet}. It summarises the means and standard deviations
(SD) of performance indexes of the inference results, whose box plots shown
in Figure~\ref{fig:arxnet-perf}.

\begin{table*}[htb]
  \centering
  \caption{Summary of inference results for ARX dynamic networks. Each statistic
    is computed from inference results of 50 random models. The label ``GIRL1''
    refers to the iterative reweighted $l_1$ method for group sparsity, ``GSBL''
    refers to Sparse Bayesian Learning and ``GSM'' refers to sampling
    methods. In each test, all methods use the same data set, except that
    ``GSM'' only uses 100 points. In the simulation of multiple SNRs, the number
    of nodes is set to 10 (i.e. $p = 10$); and in the case of multiple \#nodes,
    the SNR is set to $10$ dB. The lambdas for ``GIRL1'' are set to
    $\lambda = 0.1, 0.1, 0.01, 0.001$ for
    $\text{SNR} = 0, 10, 20, 40 \text{dB}$, respectively; and
    $\lambda = 0.05, 0.1, 0.1, 0.1$ for $p = 5, 10, 15, 20$.}

  \setlength{\tabcolsep}{10pt}
  \begin{tabular}{clrrrr}
    \toprule
    \multicolumn{2}{c}{\multirow{2}{*}{(mean$\pm$SD)}}
    & \multicolumn{4}{c}{SNRs (dB)} \\
    \cline{3-6}
    & & \multicolumn{1}{c}{0}  & \multicolumn{1}{c}{10}
      & \multicolumn{1}{c}{20} & \multicolumn{1}{c}{40} \\
    \hline
    \multirow{3}{*}{Prec (\%)}
    & {GIRL1}
    & $98.96 \pm 2.77$ & $99.58 \pm 1.73$ & $97.94 \pm 3.25$ & $97.34 \pm 4.25$
    \\
    \cline{2-6}
    & {GSBL}
    & $89.92 \pm 8.16$ & $93.05 \pm 6.95$ & $93.31 \pm 7.25$ & $93.35 \pm 6.64$
    \\ \cline{2-6}
    & {GSMC}
    & $99.89 \pm 0.74$ & $99.56 \pm 1.53$ & $99.65 \pm 1.39$ & $97.53 \pm 3.80$
    \\
    \hline
    \multirow{3}{*}{TPR (\%)}
    & {GIRL1}
    & $88.05 \pm 9.51$ & $83.06 \pm 10.40$ & $92.95 \pm 7.24$ & $95.52 \pm 5.32$
    \\
    \cline{2-6}
    & {GSBL}
    & $89.29 \pm 8.25$ & $89.14 \pm  8.45$ & $89.92 \pm 8.81$ & $88.79 \pm 8.36$
    \\
    \cline{2-6}
    & {GSMC}
    & $84.84 \pm 10.37$ & $87.80 \pm 7.50$ & $90.55 \pm 7.49$ & $94.01 \pm 6.35$
    \\ \midrule
    \multicolumn{2}{c}{\multirow{2}{*}{(mean$\pm$SD)}}
    & \multicolumn{4}{c}{\#nodes} \\
    \cline{3-6}
    & & \multicolumn{1}{c}{5}  & \multicolumn{1}{c}{10}
      & \multicolumn{1}{c}{15} & \multicolumn{1}{c}{20} \\
    \hline
    \multirow{3}{*}{Prec (\%)}
    & {GIRL1}
    & $100.00 \pm 0.00$ & $99.80 \pm 1.41$ & $99.45 \pm 1.23$ & $99.17 \pm 1.74$
    \\
    \cline{2-6}
    & {GSBL}
    & $ 94.47 \pm 9.31$ & $91.55 \pm 6.87$ & $93.66 \pm 3.57$ & $93.86 \pm 4.45$
    \\
    \cline{2-6}
    & {GSMC}
    & $100.00 \pm 0.00$ & $99.25 \pm 2.07$ & $99.01 \pm 1.97$ & $98.80 \pm 3.02$
    \\
    \hline
    \multirow{3}{*}{TPR (\%)}
    & {GIRL1}
    &$93.60 \pm 10.25$ & $86.23 \pm 8.85$ & $85.22 \pm 9.85$ & $86.82 \pm 5.41$
    \\
    \cline{2-6}
    & {GSBL}
    &$92.40 \pm 11.35$ & $89.05 \pm 8.39$ & $90.51 \pm 8.59$ & $90.89 \pm 5.22$
    \\
    \cline{2-6}
    & {GSMC}
    & $93.20 \pm 10.39$ & $88.68 \pm 8.17$ & $88.35 \pm 8.78$ & $89.67 \pm 5.05$
    \\
    \bottomrule
  \end{tabular}
  \label{tbl:arx-dynet}
\end{table*}

\begin{figure}[htbp]
  \centering

  \begin{subfigure}[b]{0.4\textwidth}
    \centering
    \includegraphics[width=\textwidth]{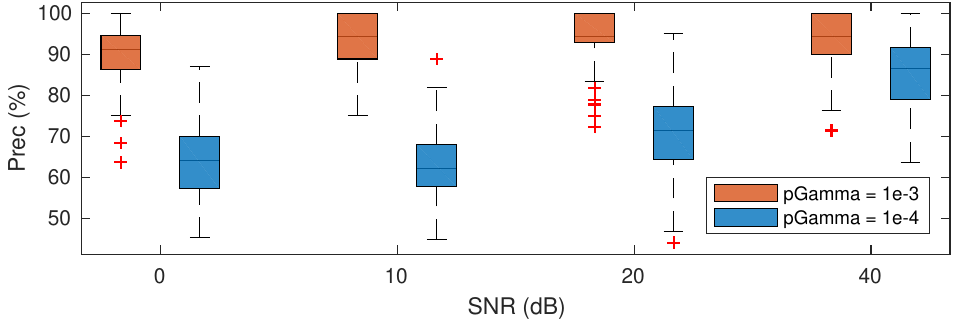}
  \end{subfigure}
  \\[5pt]
  \begin{subfigure}[b]{0.4\textwidth}
    \centering
    \includegraphics[width=\textwidth]{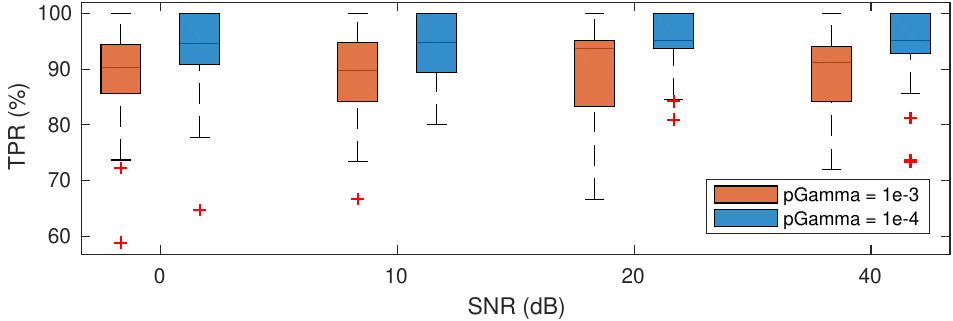}
  \end{subfigure}

  \caption{Performance of network inference using GSBL when setting different
    values to prune $\gamma$'s during the EM iterations.}
  \label{fig:SBL-perf-pGamma}
\end{figure}

One practical detail on implementing GSBL deserves our attention.  A threshold
in GSBL actually needs to be tuned to have better performance, which prunes the
elements of $\pmb{\gamma}$ (labeled as ``pGamma'' in
Fig.~\ref{fig:SBL-perf-pGamma}) in the EM iterations. It determines whether the
parameters that specific $\lambda$ corresponds to should be zero
confidently. When dealing noisy data, a larger value of ``pGamma'' works better.
The performance of different ``pGamma'' shows in Fig.~\ref{fig:SBL-perf-pGamma},
where the results with ill-selected ``pGamma'' comply with our intuition,
that is a larger SNR gives obviously better performance.

\section{An illustrative example}
\label{appdix:example}

\begin{figure}[htbp]
  \centering

  \begin{subfigure}[b]{0.2\textwidth}
    \centering
    \includegraphics[width=\textwidth]{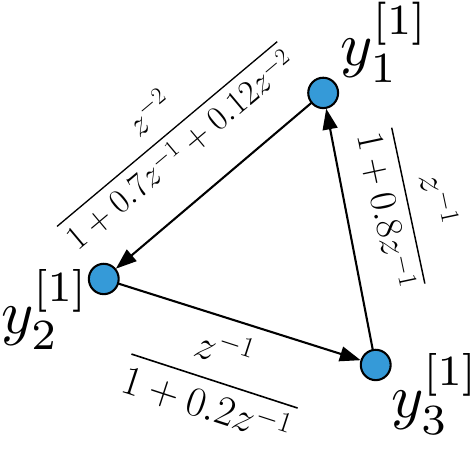}
  \end{subfigure}
  \qquad
  \begin{subfigure}[b]{0.2\textwidth}
    \centering
    \includegraphics[width=\textwidth]{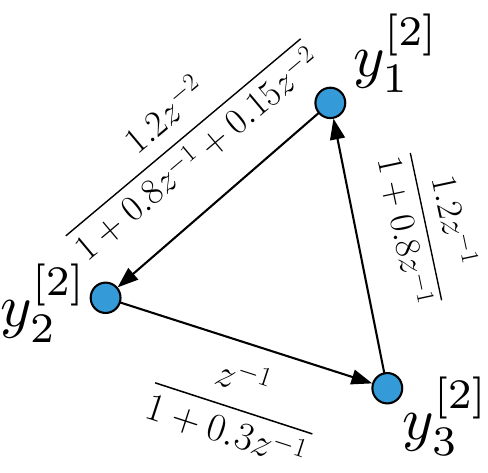}
  \end{subfigure}
  \caption{Models of the underlying system in two experiments, which share the
    same network structure but differ in parameters. Each variable is subject to
    a Gaussian white noise, which is omitted in the figure.}
  \label{fig:example-model}
\end{figure}

This appendix presents a demo example to illustrate the whole idea/setup
from Section~\ref{sec:problem-description}-~\ref{sec:heterogen-data}. In
this demo project, we perform two experiments to collect data to infer the
regulation relations between three variables $y_1, y_2$ and $y_3$. Due to
intrinsic difference of individuals or uncontrolled disturbance in
experiments, the underlying system keeps the same mechanism (i.e. the
regulation relations) but might differs in certain model parameters in two
experiments, as shown in Fig.~\ref{fig:example-model}.
The network models in Fig.~\ref{fig:example-model} can be expressed by DSFs,
\begin{equation}
  \label{eq:appdix-dsf}
  \begin{array}{ll}
    y^{[1]}(t) & = Q^{[1]}(z) y^{[1]}(t) + H^{[1]}(z) e^{[1]}(t), \\
    y^{[2]}(t) & = Q^{[2]}(z) y^{[2]}(t) + H^{[2]}(z) e^{[2]}(t),
  \end{array}
\end{equation}
where
\begin{equation}
  \label{eq:appdix-Qs}
  \begin{array}{ll}
  Q^{[1]} &=
  \begin{bmatrix}
    0 & 0 & \frac{z^{-1}}{1+0.8z^{-1}} \\
    \frac{z^{-2}}{1+0.7z^{-1}+0.12z^{-2}} & 0 & 0\\
    0 & \frac{z^{-1}}{1+0.2z^{-1}} & 0
  \end{bmatrix} \\
  Q^{[2]} &=
  \begin{bmatrix}
    0 & 0 & \frac{1.2z^{-1}}{1+0.8z^{-1}} \\
    \frac{1.2z^{-2}}{1+0.8z^{-1}+0.15z^{-2}} & 0 & 0\\
    0 & \frac{z^{-1}}{1+0.3z^{-1}} & 0
  \end{bmatrix}
  \end{array}
\end{equation}
and $H$ will be specified for convenience later on.  Now consider $y_2$ as the
example to show parametrization and the setup of multiple experiment data, where
the object is to infer that $y_1$ regulates $y_2$ but $y_3$ does not.  The whole
network can then be reconstructed by independently repeating the same procedure
on $y_1$ and $y_3$.

\begin{figure}[htbp]   
  \centering
  \includegraphics[width=.45\textwidth]{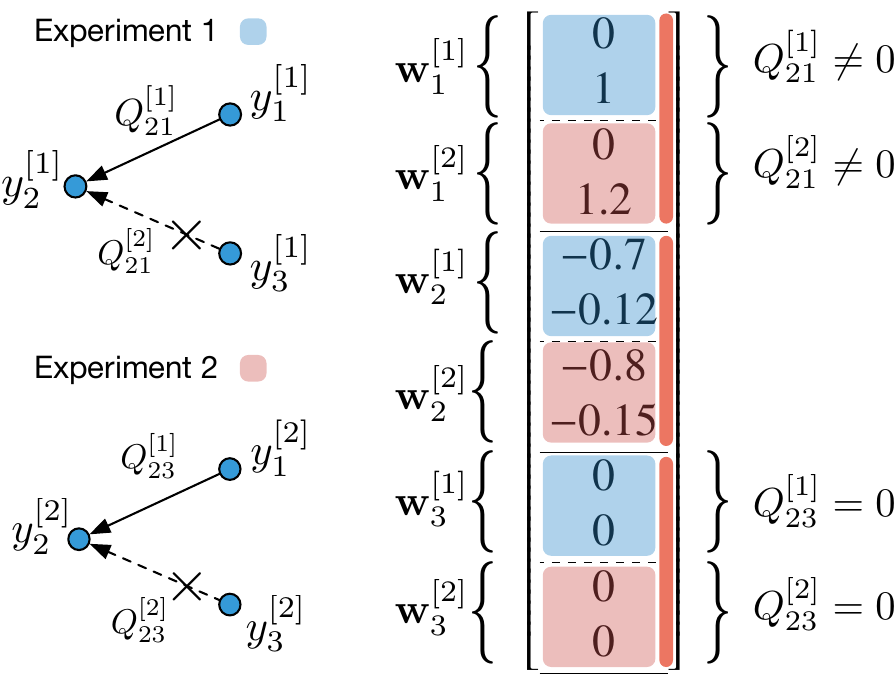}
  \caption{Illustration of parameter grouping and corresponding functions.}
  \label{fig:example-wpara}
\end{figure}

For simplicity, let this demo DSF be perfectly parametrized by ARX, and $H$
therefore is specified correspondingly,
\begin{equation}
  \label{eq:appdix-arx}
  \begin{array}{lll}
    &0.12 y_2^{[1]}(t-2) + &0.7 y_2^{[1]}(t-1) + y_2^{[1]}(t) = \\
    &&y_1^{[1]}(t-2) + e_2^{[1]}(t), \\
    &0.15 y_2^{[2]}(t-2) + &0.8 y_2^{[2]}(t-1) + y_2^{[2]}(t) = \\
    &&1.2y_1^{[2]}(t-2) + e_2^{[2]}(t). \\
  \end{array}
\end{equation}
Assuming we set all orders of polynomials in ARX to be $2$, the regression model
\eqref{eq:pseudolinear-regression-form} is then be expressed as following
\begin{equation}
  \label{eq:appdix-regression}
  \begin{array}{lll}
    \hat{y}_2^{[1]}(t) &=&
      \begin{array}[t]{llllll}
        [ & y_1^{[1]}(t-1) & y_1^{[1]}(t-2) & y_2^{[1]}(t-1) & y_2^{[1]}(t-2) & \\
          & y_3^{[1]}(t-1) & y_3^{[1]}(t-2) & ]
      \end{array}\\
    &&\big[
      \begin{array}{cc:cc:cc}
        0 & 1 & -0.7 & -0.12 & 0 & 0
      \end{array}
      \big]^T, \\
    \hat{y}_2^{[2]}(t) &=&
      \begin{array}[t]{llllll}
        [ & y_1^{[2]}(t-1) & y_1^{[2]}(t-2) & y_2^{[2]}(t-1) & y_2^{[2]}(t-2) & \\
          & y_3^{[2]}(t-1) & y_3^{[2]}(t-2) & ]
      \end{array}\\
    &&\big[
      \begin{array}{cc:cc:cc}
        0 & 1.2 & -0.8 & -0.15 & 0 & 0
      \end{array}
      \big]^T. \\
  \end{array}
\end{equation}
Consider time points $t_1, t_2, t_3$ and set up the measurements from two
experiments as \eqref{eq:regression-form-exp-all-N-blocks} in
Section~\ref{sec:heterogen-data}, yielding \eqref{eq:appdix-setup-data}.
Now we investigate the grouping patterns in the vector of parameters, which are
marked in different colours in Fig.~\ref{fig:example-wpara}.
The parameters marked by blue boxes are ones of the model in Experiment~1 and
ones marked by red boxes are of Experiment~2.  Considering the consistency of
network structures over experiments, we are obliged to guarantee the blue and
red subgroups in one group indicated by orange boxes are both zero or nonzero.
With the rearrangement of regressors as \eqref{eq:regression-form-exp-all}, the
first and third groups determine which links exist in all experiment. To be
specific, the arc from $y_1$ to $y_2$ should exist since both $Q_{21}^{[1]}$ and
$Q_{21}^{[2]}$ are nonzero, while there is no arc from $y_3$ to $y_2$ since both
$Q_{23}^{[1]}$ and $Q_{23}^{[2]}$ are identical to zero. Hence, in
identification, we perform group sparsity with respect to groups indicated by
orange boxes, which meanwhile also guarantees sparsity of network structures.

\begin{sidewaysfigure*}
  \centering
  \begin{equation}
    \label{eq:appdix-setup-data}
    \begin{array}{lll}
    \left[
    \begin{array}{c}
      y_2^{[1]}(t_1) \\ y_2^{[1]}(t_2) \\ y_2^{[1]}(t_3) \\ \hdashline
      y_2^{[2]}(t_1) \\ y_2^{[2]}(t_2) \\ y_2^{[2]}(t_3)
    \end{array}
    \right] &=&
    \left[
      \begin{array}{cccc|cccc|cccc}
        y_1^{[1]}(t_1-1)  & y_1^{[1]}(t_1-2) & 0 & 0 &
        y_2^{[1]}(t_1-1)  & y_2^{[1]}(t_1-2) & 0 & 0 &
        y_3^{[1]}(t_1-1)  & y_3^{[1]}(t_1-2) & 0 & 0 \\
        y_1^{[1]}(t_2-1)  & y_1^{[1]}(t_2-2) & 0 & 0 &
        y_2^{[1]}(t_2-1)  & y_2^{[1]}(t_2-2) & 0 & 0 &
        y_3^{[1]}(t_2-1)  & y_3^{[1]}(t_2-2) & 0 & 0 \\
        y_1^{[1]}(t_3-1)  & y_1^{[1]}(t_3-2) & 0 & 0 &
        y_2^{[1]}(t_3-1)  & y_2^{[1]}(t_3-2) & 0 & 0 &
        y_3^{[1]}(t_3-1)  & y_3^{[1]}(t_3-2) & 0 & 0 \\ \hdashline
        0 & 0 & y_1^{[2]}(t_1-1)  & y_1^{[2]}(t_1-2) &
        0 & 0 & y_2^{[2]}(t_1-1)  & y_2^{[2]}(t_1-2) &
        0 & 0 & y_3^{[2]}(t_1-1)  & y_3^{[2]}(t_1-2) \\
        0 & 0 & y_1^{[2]}(t_2-1)  & y_1^{[2]}(t_2-2) &
        0 & 0 & y_2^{[2]}(t_2-1)  & y_2^{[2]}(t_2-2) &
        0 & 0 & y_3^{[2]}(t_2-1)  & y_3^{[2]}(t_2-2) \\
        0 & 0 & y_1^{[2]}(t_3-1)  & y_1^{[2]}(t_3-2) &
        0 & 0 & y_2^{[2]}(t_3-1)  & y_2^{[2]}(t_3-2) &
        0 & 0 & y_3^{[2]}(t_3-1)  & y_3^{[2]}(t_3-2)
      \end{array}
    \right] \\
    &&
    \left[
      \begin{array}{c}
        0 \\ 1 \\ \hdashline 0 \\ 1.2 \\ \hline
        -0.7 \\ -0.12 \\ \hdashline -0.8 \\ -0.15 \\ \hline
        0 \\ 0 \\ \hdashline 0 \\ 0
      \end{array}
    \right] +
    \left[
      \begin{array}{c}
        e_2^{[1]}(t_1) \\ e_2^{[1]}(t_2) \\ e_2^{[1]}(t_3) \\ \hdashline
        e_2^{[2]}(t_1) \\ e_2^{[2]}(t_2) \\ e_2^{[2]}(t_3)
      \end{array}
    \right]
    \end{array}
  \end{equation}
\end{sidewaysfigure*}

\end{document}

%% file: supports/userdef-mathsym.tex
\newcommand{\mnorm}[1]{{\left\vert\kern-0.25ex\left\vert\kern-0.25ex\left\vert #1
    \right\vert\kern-0.25ex\right\vert\kern-0.25ex\right\vert}} 
\newcommand{\bm}[1]{\begin{bmatrix} #1 \end{bmatrix}}

\newcommand{\diag}{\operatorname{{diag}}}

\newcommand{\minimize}[1]{\underset{#1}{\operatorname{minimize}\;}}  

\newcommand{\argmax}[1]{\underset{#1}{\operatorname{arg\,\!max}\;}}
\newcommand{\argmini}[1]{\operatorname{arg}\,\!\underset{#1}{\operatorname{min}}\;}
\newcommand{\argmaxi}[1]{\operatorname{arg}\,\!\underset{#1}{\operatorname{max}}\;}

